# Disentangling forced and internal components in observed climate evolution using spatiotemporal deep learning

Parvathi Kooloth[1], Wenyu Zhou[1*], L. Ruby Leung[1],

Shang-Ping Xie[2], Hailong Wang[1], and Qiang Fu[3]

[1]Pacific Northwest National Laboratory, Richland, WA, USA 99352

[2]University of California, San Diego, La Jolla, CA, USA 92093

[3]University of Washington, Seattle, WA, USA 98195

Corresponding Author: wenyu.zhou@pnnl.gov

## Abstract

Attribution of observed historical climate evolution to externally forced change and internal variability is fundamental to understanding climate change and evaluating climate models. Yet such attribution remains challenging because the forced response is intertwined with internal variability in observations. Here, we develop a convolutional long short-term memory network (ConvLSTM) to disentangle these two components. Trained and evaluated using large-ensemble simulations from multiple climate models, ConvLSTM consistently outperforms existing model-based, statistical, and machine-learning approaches. Applying ConvLSTM to observations reveals an evolving spatial pattern of the forced response and substantial contributions from internal variability to recent multidecadal climate trends. For several key climate features whose attribution remains debated, ConvLSTM indicates that internal variability amplified the forced reversal of Atlantic interhemispheric temperature asymmetry, dominated the observed La Niña-like warming pattern over the equatorial Pacific and the associated precipitation decline over southwestern North America since 1980, and modulated Arctic amplification around its nearly constant threefold forced value. Comparison between the ConvLSTM-estimated and model-simulated forced responses reveals broadly consistent patterns but also important discrepancies, with climate models simulating stronger Northern Hemisphere warming, weaker subtropical high intensification, and smaller land precipitation declines. Overall, ConvLSTM provides a robust framework for attributing historical climate evolution and evaluating the fidelity of climate models.

## Introduction

The Earth's climate, influenced by both external forcing and internal variability, has undergone substantial change and variability since 1950 (Vecchi et al. 2006; Hu and Fu 2007; Ting et al. 2009; Deser et al. 2010; Kosaka and Xie 2013; Dai et al. 2015; Xie et al. 2015; Masson-Delmotte et al. 2021; Rantanen et al. 2022). A central question is how much of the observed historical trend is due to external forcings and how much is due to internal variability (Otterå et al. 2010; Baek et al. 2022; He et al. 2023; Hua and Dai 2024; Klavans et al. 2025; Xie et al. 2026). Accurate attribution is critical not only for understanding climate change but also for evaluating climate models (e.g., Seager et al. 2019; Watanabe et al. 2021; Wills et al. 2022; Sweeney et al., 2023; Sweeney et al., 2024; Zhou, Leung, and Lu 2024; Simpson et al. 2025; Zhou et al. 2026). Such attribution, however, remains challenging because reality provides a single realization in which the forced and internal components are not readily separable. The forced response, driven by multiple forcing agents including greenhouse gases, aerosols, and land-use changes, is evolving both temporally and spatially (e.g., Shindell 2014). Internal variability, characterized by various modes such as the El Niño–Southern Oscillation (ENSO) (Philander 1983; Timmermann et al. 2018), Interdecadal Pacific Oscillation (IPO) (Newman et al. 2016), and Atlantic Multidecadal Variability (AMV) (Zhang et al. 2019), encompasses a range of spatiotemporal evolutions that are not necessarily orthogonal to the forced response in their spatial pattern and temporal evolution (e.g., Corti, Molteni, and Palmer 1999).

Various model-based, statistical, and machine-learning (ML) approaches have been proposed to disentangle the forced response and internal variability. Unlike the single realization of observations, climate models can create a large ensemble of simulations under the same time-evolving forcing (e.g., Eyring et al. 2016; Deser et al. 2020). The mean of the large ensemble runs averages out internal variability and yields an estimate of the forced response. However, climate models exhibit biases, which limit our confidence in their direct estimates of the forced response. The limitations of climate models motivate the development of data-driven approaches to identify the forced response from observations. The simplest approach is to use the long-term linear trend.

But low-frequency internal variability—such as IPO and AMV—projects strongly onto multidecadal trends so that even trends over several decades do not accurately reflect the forced response. More sophisticated approaches, including empirical orthogonal function (EOF), low-frequency filtering, and optimal fingerprinting, exploit statistical regularities to distinguish slowly evolving, spatially coherent signals from higher-frequency variability. However, many of these methods implicitly assume fixed spatial patterns of the forced response or rely on strong linearity assumptions. In reality, the forced response evolves in both pattern and magnitude over time, due to changes in forcings and feedbacks. Recent community efforts, particularly the Forced Component Estimation Statistical Method Intercomparison Project (ForceSMIP) (Wills et al. 2026), systematically evaluate the existing model, statistical, and ML methods based on the benchmark provided by the large-ensemble model simulations. Some methods show skill in extracting the forced response, but the skill is often inconsistent across variables, and the root-mean-square-errors (RMSEs) remain substantial. Moreover, the investigated ML methods present skill comparable to, if not worse than, the EOF- or fingerprint-based statistical methods. These results highlight the promise and unfulfilled potential in data-driven approaches for forced-response estimation.

In this study, we develop a convolutional long short-term memory network (ConvLSTM) to disentangle the forced and internal components in observations. ConvLSTM combines convolutional layers that capture spatial structure with recurrent layers that learn temporal dependencies, making it well-suited for the attribution problem here. We first describe the ConvLSTM architecture and its training and evaluation using large-ensemble simulations of climate models. After demonstrating the skill of ConvLSTM, we apply ConvLSTM to observations, revealing a spatially and temporally varying forced response and substantial contributions from internal variability. We further quantify the contributions of the forced and internal components to historical evolutions in several key climate features and discuss the ability of climate models in capturing the forced response. Overall, this study presents ConvLSTM as a powerful method for climate attribution and provides important insights into historical climate change and internal variability.

## ConvLSTM: Architecture, training and evaluation

The Convolutional LSTM (Shi et al. 2015) extends the long short-term memory model (Hochreiter and Schmidhuber 1997) by replacing fully connected transformations with convolutional operations (Lecun et al. 1998), thereby preserving spatial structure while handling temporal dependencies. The key advantage of LSTM over standard autoregressive models lies in its gated architecture: the input gate extracts relevant features from the current input and hidden state, while the forget gate selectively retains slowly evolving features and discards those with shorter memory (see a schematic in Fig. 1). The combination of LSTM and convolutional operations allows ConvLSTM to jointly learn spatial and temporal features, making it well suited for disentangling spatiotemporal evolutions.

The learning task for ConvLSTM here is to decompose observed historical climate evolution into the forced and internal components. For this task, ConvLSTM is trained and evaluated using the large-ensemble simulations of multiple climate models (Table S1), in which the forced response is known from the ensemble mean, whereas the internal variability is the difference between individual realizations and the ensemble mean. For training, the input is the anomaly fields of surface temperature (Ts), precipitation (Pr), and sea-level pressure (SLP) in individual realizations and the output target is the internal variability in individual realizations, that is,

$$x_{i,\alpha}(t) \xrightarrow{ConvLSTM} x_{i,\alpha}^{iv}(t) = x_{i,\alpha}(t) - \overline{x_{\alpha}}(t), \qquad (1)$$

where $x$ represents the anomalies relative to the climatology, $\alpha$ represents the model, $i$ represents the realization, $iv$ represents the internal variability, and overline indicates the true forced response estimated using the ensemble mean. Here, we choose internal variability rather than the forced response as the output target, because the forced response is identical across all realizations for a given model and therefore makes ConvLSTM more prone to overfitting. ConvLSTM can be viewed as a nonlinear autoregressive model that learns the evolution of internal variability based on its current input and preceding hidden and cell states (Fig. 1). With internal variability learned, the forced component is derived as,

$$x_{i,\alpha}^{f}(t)_{ConvLSTM} = x_{i,\alpha}(t) - x_{i,\alpha}^{iv}(t)_{ConvLSTM}. \qquad (2)$$

Performance is evaluated based on the unseen model data (Table S1) by comparing the estimated forced response $x^{f}_{i,\alpha}(t)_{ConvLSTM}$ with the true forced response $\overline{x_{\alpha}}(t)$ derived from the ensemble mean. We compare ConvLSTM against six other methods: the multi-model mean (MMM), the leading EOF (EOF-1st), the low-frequency component analysis (LFCA) (Wills et al. 2020), the warming-scaled MMM (TrainingEM), EOF-smoothed linear regression (EOF-SLR) (Gavrilov et al. 2024), and the ML-based UNet method (UNet) (Bône et al. 2024). The first two are widely used baseline methods whereas the latter four are selected as representative methods from ForceSMIP (Methods). For a fair comparison with the ForceSMIP methods, we evaluate the seven methods using the same five evaluation cases as ForceSMIP. Because five evaluation cases alone may be insufficient for a statistically robust assessment, we additionally evaluate ConvLSTM, MMM, and EOF-1st using multiple realizations from the five evaluation models, rather than a single realization for each model (Table S1).

We begin by evaluating the skill of different methods in extracting the forced response of Ts (Fig. 2). Their performance is first illustrated using evaluation case 1B (ACCESS-ESM1-5 r1), the same illustration case adopted in ForceSMIP. Evident in this evaluation case, the full trend over 1980-2022 (Fig. 2A) differs considerably from the true forced response (Fig. 2B), highlighting the influence of internal variability on the multidecadal trend. The skill of the attribution methods thus depends on how they isolate the internal variability and uncover the forced response. EOF-1st and LFCA estimate a forced trend that largely resembles the full trend, indicating low skill in isolating the internal variability; MMM, TrainingEM, EOF-SLR, and UNet capture the large-scale pattern of the forced trend but exhibit notable biases in regional features, particularly over the tropical Pacific; ConvLSTM performs well in capturing regional features such as the enhanced warming over the east equatorial Pacific and the muted warming over the subpolar North Atlantic and Southern Ocean (Fig. 2C-I). We then quantify the skills of different methods based on the global mean (area weighted) RMSE of the estimated forced response relative to the true forced response. According to the time series of the median RMSE across the five evaluation cases, ConvLSTM outperforms the other methods over most of the historical period (Fig. 2J). EOF-1st yields the largest RMSE, likely because it assumes a fixed spatial pattern of the forced response; MMM, TrainingEM, and EOF-SLR perform better than LFCA and UNet but have RMSE higher than ConvLSTM. Furthermore, according to the time mean RMSE of the forced response (Fig. 2K) and the RMSE of the forced trend over 1980-2022 (Fig. 2L), ConvLSTM outperforms other methods for evaluation models of ACCESS-ESM1-5, GFDL-SPEAR, IPSL-CM6A-LR, and NorCPM1, based on either single (as ForceSMIP) or multiple realizations. The second best-performing method varies across evaluation models, indicating that no existing method consistently provides reliable and robust estimates as ConvLSTM.

The evaluation case EC-Earth3 r6 is an exception where ConvLSTM performs worse than several other methods, but ConvLSTM does perform better when evaluated using multiple realizations of EC-Earth3. Further analysis indicates that EC-Earth3 r6 is an extreme outlier associated with unrealistic Atlantic Meridional Overturning Circulation (AMOC) variability and therefore may not represent a realistic analogue of the historical record (Fig. S1).

We next examine the method skill in extracting the forced response in Pr (Fig. 3). Again, visualized based on the evaluation case 1E, the full trend (Fig. 3A) differs substantially from the true forced trend (Fig. 3B), highlighting the effect of internal variability. LFCA and UNet estimate a forced trend that largely resembles the full trend; EOF-1st and EOF-SLR capture the large-scale pattern over the ocean but fail to capture the wetting trends over the Sahel and high latitudes; MMM and TrainingEM perform better but still exhibit notable biases in regional changes over the equatorial Pacific; ConvLSTM appears to provide the closest estimate of the forced trend pattern (Fig. 3C-I). Quantitatively, according to the RMSEs of the forced response (Fig. 3J,K) and of the forced trend over 1980-2022 (Fig. 3L), ConvLSTM robustly outperforms other methods.

Consistently, ConvLSTM demonstrates better skill in extracting the forced response in sea-level pressure (Fig. S2). Together, these results demonstrate that ConvLSTM provides a consistently more accurate estimate of the forced response across variables and climate models, establishing a robust foundation for observational attribution.

**Attribution of the observed trends to the forced and internal components**

Having demonstrated its strong skill on unseen model data, we now apply ConvLSTM to attributing observed historical climate evolution to the forced and internal components. We develop a suite of ten ConvLSTM models, each trained on a distinct subset of ten climate models drawn from the full set of 15 models (Table S1). This ensemble framework enables quantification of the inferential uncertainty and reduces biases associated with individual climate models. To highlight the temporally evolving nature of the forced response and internal variability, we investigate the trends over two periods, 1980-2022 and 1950-1992.

Over 1980-2022, the observed Ts trend presents enhanced warming in the west equatorial-to-midlatitudinal Pacific, slight cooling in the subtropical southeast Pacific and Southern Ocean, enhanced warming in the Arctic and muted warming over the high-latitudinal Eurasia and North America (Fig. 4A). The observed SLP trend features a positive trend over tropical east Pacific that strengthens the Walker cell and dipolar changes over the high latitudes that correspond to intensification and poleward expansion in subtropical highs (Fig. 4B). The observed Pr trend shows intensified ITCZ and drying over the subtropical southwest ocean and adjacent land regions (e.g., U.S. Southwest) located equatorward of the enhanced SLP (Fig. 4C). The forced response estimated by ConvLSTM indicates that the observed trends are partially contributed by the forced response (Fig. 4G-I vs Fig. 4A-C). In particular, the forced response contributes to the enhanced warming over the mid-latitudinal west Pacific and muted warming over the subtropical southeast Pacific and Southern Ocean (Fig. 4G), intensification and poleward expansion in subtropical highs (Fig. 4H), and the subtropical drying associated with the poleward expansion (Fig. 4I). Nevertheless, substantial discrepancies exist between the observed and forced trend, highlighting the contribution of internal variability (Fig. 4M-O). Specifically, internal variability contributes to a negative IPO-like pattern with cooling over the equatorial to subtropical east Pacific (Fig. 4M), intensified SLP over the tropical-to-mid-latitudinal east Pacific (Fig. 4N), and the associated drying over the equatorial Pacific and the subtropics (Fig. 4O). Also, internal variability contributes to cooling over the high-latitudinal Eurasia and North America, as well as cooling over the Southern Ocean (Fig. 4M).

The observed trends over 1950-1992 are distinct from those over 1980-2022 (Fig. 4D-F vs. Fig. 4A-C). In particular, the 1950-1992 trends feature cooling over the mid-latitudinal west Pacific (Fig. 4D), enhanced warming and negative SLP trends over the tropical east Pacific (Fig. 4E), wetting over the U.S. Southwest but drying over the tropical Atlantic and Sahel (Fig. 4F). The distinct trends between the two periods are contributed by both the forced response (Fig. 4J-L) and the internal variability (Fig. 4P-R). In contrast to 1980-2022, a period characterized by rapidly increasing CO2 concentration and regionally dependent change in aerosol emission, the period of 1950-1992 features a slowly increasing CO2 concentration and consistent increase in aerosol emission. Expected from these forcing differences, the forced trend over 1950-1992 features weak global warming but an inter-hemispheric asymmetric pattern with slight cooling in the Northern Hemisphere (Fig. 4J), slight weakening of the North Pacific high (Fig. 4K), and a cross-equator southward shift in tropical precipitation associated with the asymmetric warming pattern (Fig. 4L). The contribution of internal variability during 1950-1992 also differs markedly from that during 1980-2022. Whereas the latter period features a negative IPO-like pattern (Fig. 4M-O), the earlier period features a positive IPO-like pattern in the Pacific Ocean and a negative AMV-like pattern in the Atlantic Ocean (Fig. 4P-R).

Overall, the ConvLSTM-based attribution highlights three key findings. First, the observed trends over a period of 42 years are partially contributed by the externally forced response but also substantially modulated by internal variability. Second, the forced response evolves over time, consistent with the changing external forcing. Third, internal variability contributes substantially to multidecadal trends, through major modes like IPO and AMV.

**Contributions of forced and internal components to historical evolutions in key climate features**

The difficulty in separating the forced and internal components has limited our understanding of historical climate change and introduced uncertainty in interpreting the discrepancy between observations and model simulations. For instance, the global warming hiatus over 1998-2012 was initially interpreted as a slowdown of the anthropogenic warming and evidence of model bias (Fyfe, Gillett, and Zwiers 2013) until later recognized to be internal variability (Meehl et al. 2011; Kosaka and Xie 2013; Watanabe et al. 2014; England et al. 2014; Dai et al. 2015). Here, we apply ConvLSTM to historical evolutions in several key climate features for which the relative contributions of forced and internal components remain debated.

We first investigate three key temperature features – the global mean temperature, the Arctic mean temperature, and the Atlantic interhemispheric asymmetry (Fig. 5). Illustrated by two evaluation cases 1E (GFDL-SPEAR r3) and 1J (NorCPM1 r4), ConvLSTM well extracts the forced responses in these features (Fig. 5A-F). First, despite the presence of strong internal variability, ConvLSTM successfully extracts the smooth evolution of the forced response, close to the true forced response. Second, ConvLSTM can capture the different forced response between two cases, i.e., the different global warming rates (Fig. 5A,D), the different ratios between the forced Arctic and global warming (Fig. 5B,E), and the distinct evolutions of the Atlantic temperature asymmetry--- swing-like evolution in GFDL-SPEAR but continuous decrease in NorCPM1 (Fig. 5C,F). The spatial difference in the forced temperature response between the two models, likely contributed by the differences in their AMOC declines, are well captured by ConvLSTM (Fig. S3). The fact that ConvLSTM can distinguish the model-dependent forced SST responses demonstrates its skill beyond the fixed estimate by MMM and enhances our confidence in applying ConvLSTM to attribute historical evolution in temperature features.

The observed global mean temperature features a long-term warming trend alongside strong interannual variability and an apparent hiatus over 1998-2012 (Fig. 5G). Compared to observations, the ConvLSTM-estimated forced response exhibits a smoother trajectory, characterized by a gradually increasing warming rate consistent with historical evolutions of greenhouse/aerosol forcing and intermittent cooling consistent with historical volcanic forcing (Fig. 5G). Internal variability (IV) estimated by ConvLSTM features interannual peaks and decadal warm and cold periods, with the warming hiatus over 1998-2012 correctly attributed to internal variability (Fig. 5J). The estimated IV allows a direct identification of internal modes from the internal component, instead of relying on approximate methods to remove the forced response from observations. EOF analysis on IV over the Pacific Ocean identifies the so-called interdecadal Pacific Oscillation (IPO) as the leading mode, which explains much of the variability in global mean Ts (Fig. 5J,M). In particular, the correlation between global Ts internal variability and IPO is r=0.68 on interannual timescale and r=0.69 on decadal timescale (Fig. 5J).

As a well-known feature of climate change, the Arctic warming is amplified relative to the global warming (Manabe and Stouffer 1980; Holland and Bitz 2003; Serreze et al. 2009). The degree of Arctic Amplification (AA), measured as the ratio between the Arctic and global mean warming, however, varies with time in historical observations. Over recent decades, with reduced global warming rate but enhanced Arctic warming rate, the degree of AA increased rapidly to four (Rantanen et al. 2022) (Fig. 5H). Recent studies suggest that this fourfold AA is positively contributed by internal variability (Sweeney et al. 2023; Zhou et al. 2024). ConvLSTM estimates a forced Arctic warming that accelerates gradually in concert with the forced global warming, yielding a nearly constant threefold AA throughout the historical period (Fig. 5H). Internal variability estimated by ConvLSTM features an anomalous warm period for the Arctic over 2005-2020 (Fig. 5K). EOF analysis on IV indicates that an Arctic mode (AM), identified as the leading EOF of IV north of 65$^o$N (Fig. 5K,N; Methods), dominates the Arctic Ts variability. The correlation between Arctic Ts internal variability and AM is r=0.85 on interannual timescale and r=0.95 on decadal time scale (Fig. 5K). Over recent decades, as IPO suppressed global warming (Fig. 5J) and AM amplified Arctic warming (Fig. 5K), the degree of AA arose from its forced value of three to the observed value of four, consistent with the conclusion of Zhou et al. (2024).

Atlantic interhemispheric temperature asymmetry exerts a profound influence on global climate and weather (Xie and Carton 2004; Knight, Folland, and Scaife 2006; Zhang et al. 2019). In particular, the suppressed Atlantic hurricane activity and the prolonged Sahel drought during the 1970s and 1980s are linked to negative anomalies in this temperature asymmetry (Goldenberg et al. 2001; Giannini, Saravanan, and Chang 2003; Zhang and Delworth 2006). However, the extent to which these anomalies are externally driven or internally generated remains debated (Otterå et al. 2010; Baek et al. 2022; He et al. 2023; Hua and Dai 2024). The observed Atlantic temperature asymmetry since 1950 exhibits a V-shaped trajectory, with an initial decline followed by a sustained increase, superimposed on substantial internal variability (Fig. 5I). Attribution using ConvLSTM indicates that the observed historical evolution is contributed by both the forced response and internal variability. Specifically, the forced response exhibits a smooth decrease followed by an increase, consistent with the historical rise and subsequent decline of Northern Hemisphere aerosol forcing; while internal variability amplifies the swing and gives rise to the pronounced negative anomalies during the 1970s and 1980s (Fig. 5I). EOF analysis indicates that internal variability in the Atlantic temperature asymmetry is dominated by the so-called Atlantic multidecadal variability (AMV), identified here as the leading EOF of IV over the Atlantic Ocean (Fig. 5L,O; Methods). The correlation between the Atlantic temperature asymmetry internal variability and AMV is r=0.35 on interannual timescale and r=0.45 on decadal timescale (Fig. 5L).

We next examine three regional climate features --- the equatorial Pacific zonal temperature gradient, the northeastern Pacific SLP, and the southwestern North American precipitation (Fig. 6; Methods). Over 1980-2022, the equatorial Pacific experienced a La Niña-like warming pattern, corresponding to an enhanced zonal temperature gradient (Fig. 6G,J). This observed La Niña-like pattern is rarely captured by model simulations, whose ensemble mean (i.e., the forced response) often presents a weak El Niño-like pattern. Whether this observation–model discrepancy reflects systematic biases in the simulated forced response (Seager et al. 2019; Wills et al. 2022) or instead arises primarily from internal variability (Watanabe et al. 2021, 2024; Sweeney et al. 2024; Xie et al. 2026) remains a subject of active debate. During the same period, observations also show pronounced drying over southwestern North America, accompanied by substantial intensification of the northeastern Pacific subtropical high (Fig. 6G,K,L). In contrast, climate models generally simulate only modest drying over southwestern North America together with much weaker intensification of the northeastern Pacific SLP (Jacobson et al. 2024; Zhou, Leung, Harrop, et al. 2026).

We first demonstrate, through evaluation case 1E, that the full trends in these features can differ substantially from the forced trends and that ConvLSTM successfully extracts the latter. The full anomaly of case 1E exhibits time-evolving trends (Fig. 6A,D-F) closely resembling observations (Fig. 6G,J-L). In contrast to the 1950-1992 period, the 1980-2022 period features a La Niña-like warming pattern, intensified northeastern Pacific SLP, and pronounced drying over the southwestern North America (Fig. 6A). The true forced response, however, features a weak El Niño-like warming pattern and much weaker SLP intensification and precipitation decline (Fig. 6B), reflecting substantial influences from internal variability (Fig. 6C). Despite the strong influence of internal variability, ConvLSTM successfully recovers the forced response, with the estimated forced response closely matching the true forced response for all three regional features (Fig. 6D-F).

Applying ConvLSTM to observations reveals that the historical evolution of all three regional climate features is strongly influenced by internal variability. The observed series exhibit strong interannual variability and opposite multidecadal trends during 1980-2022 and 1950-1992 (Fig. 6J-L). In contrast to the observed trend (Fig. 6G), the ConvLSTM-estimated forced trends over 1980-2022 feature a weak El Niño-like warming pattern that weakens the equatorial Pacific temperature gradient (Fig. 6H,J), only modest intensification of the northeastern Pacific SLP (Fig. 6H,K), and a slight decline in the southwestern North America precipitation (Fig. 6H,L). These results indicate that the observed trends over 1980-2022 are dominated by internal variability (Fig. 6I), which reversed the weak forced El Niño-like pattern into a La Niña-like pattern and substantially enhanced the regional SLP intensification and precipitation decline (Fig. 6J-L). It turns out that internal variability of all three features is regulated by IPO (Fig. 6M-O). IPO is correlated with the equatorial Pacific temperature gradient, northeastern Pacific SLP, and southwestern North American precipitation, respectively with interannual correlations of $r = 0.65$, $-0.58$, and $0.43$, and even stronger decadal correlations of $r = 0.67$, $-0.79$, and $0.7$.

**Comparison of the forced responses estimated by ConvLSTM and climate models**

Without an explicit estimate of the forced response, model biases have been inferred by comparing ensemble runs of models with the single realization of observations. When model ensemble runs rarely reproduce the observed trend, it is often interpreted as evidence of systematic biases in the simulated forced response. However, such an interpretation depends critically on how rare the observed sequence is and how well the models represent internal variability. If the observed sequence is strongly influenced by internal variability and therefore genuinely rare, or if models systematically underrepresent internal variability, the statistical fact that model runs rarely reproduce the observed trend does not mean a definitive model bias. By providing an explicit estimate of the forced response, ConvLSTM enables a direct assessment of the forced response simulated by climate models.

We first compare the spatial patterns of the forced trends over 1980-2022 estimated by ConvLSTM (Fig. 7A-C) and by climate models (Fig. 7D-F). Overall, the two estimates are broadly consistent, sharing prominent features including enhanced warming over land, the Kuroshio Extension, and high latitudes; reduced warming over the subpolar north Atlantic and Southern Ocean; intensified and poleward-shifted subtropical highs; increased precipitation over the equatorial and high latitudes but decreased precipitation over the subtropics. Nevertheless, certain differences exist between the forced trends estimated by models and ConvLSTM (Fig. 7G-I; stipples denote regions at least 10 out of 15 models agree on the sign). Relative to ConvLSTM, climate models simulate stronger warming over the Northern Hemisphere, weaker SLP intensification over the subtropics, and excessive precipitation increase over the central equatorial Pacific, subtropical land regions, and the Sahel. Such differences are consistently seen if the forced trends estimated by models are normalized by their respective global mean warming rates before constructing MMM (Fig. S4).

We further compare climate models and ConvLSTM on several key climate features. For the global and arctic mean temperature, MMM simulates higher warming rates than ConvLSTM, although the differences lie within the inter-model spread (Fig. 7J,K). For the degree of AA, MMM simulates a forced threefold AA, largely consistent with ConvLSTM (Fig. 7L). For the interhemispheric temperature asymmetry, MMM simulates a similar swing but a larger long-term increasing trend than ConvLSTM, implying a stronger response to greenhouse gases in climate models, although the difference again falls within the inter-model spread (Fig. 7M). For the equatorial Pacific zonal temperature gradient, MMM simulates a weak forced El-Niño-like pattern similar to ConvLSTM (Fig. 7N). For the SLP intensification over the northeast Pacific and over the zonal subtropics, MMM simulates a weaker intensification and the difference in the zonal subtropics is shared by most models (Fig. 7H,O,P). For the precipitation decline over the southwestern North America and over the global land, climate models simulate weaker declines. The weaker precipitation declines are dynamically consistent with the weaker subtropical SLP intensification (Fig. 7H,I) and shared by most models (Fig. 7I,Q,R). This suggests that climate models may underestimate land precipitation and humidity declines under climate change.

Finally, for several key climate features – including Arctic Amplification, equatorial Pacific zonal Ts gradient, northeast Pacific SLP, zonal subtropical zonal SLP, and southwestern North America and global land Pr— the observed trends over 1980-2022 lie outside of the uncertainty range of the estimated forced response (Fig. 7J-R). This again demonstrates that internal variability exerts a substantial influence on observed multidecadal climate trends and should be explicitly accounted for when interpreting observation–model discrepancies.

**Summary and Discussion**

A fundamental challenge in climate science is to determine how much of the observed historical climate evolution arises from externally forced climate change versus internally generated climate variability. Resolving this question is essential for climate attribution, interpretation of observed trends, and evaluation of climate models, yet it remains difficult because observations provide only a single realization in which forced and internal signals are intertwined. In this study, we developed a ConvLSTM framework to separate the forced and internal signals in the historical climate evolution. Comprehensive evaluation against the ForceSMIP benchmark demonstrates that ConvLSTM consistently outperforms existing model-based, statistical, and machine-learning methods across multiple climate variables. It reproduces the evolving pattern of the forced response and distinguishes the distinct forced responses among different climate models. The resulting improvement in forced-response estimation provides a more robust basis for attributing historical climate evolution and assessing the fidelity of climate model simulations.

Applying ConvLSTM to observations reveals a temporally evolving forced response and substantial effects of major internal climate modes. Specifically, the forced response presents a distinct spatial pattern between the periods of 1950-1992 and 1980-2022, consistent with the distinct historical forcings. Internal climate modes, including IPO, AMV, and AM, are robustly identified from the ConvLSTM-estimated internal variability and are shown to substantially modulate observed multidecadal climate trends. Over recent decades, internal variability reduced the global mean warming but enhanced the Arctic warming, amplified the forced swing in the Atlantic temperature asymmetry, reversed the weak forced El Niño-like warming pattern into the observed La Niña-like pattern, and intensified the subtropical sea-level pressure intensification and the associated land precipitation decline.

The ConvLSTM-estimated forced response enables a direct assessment of the forced response simulated by climate models. Overall, the estimated spatial patterns are broadly consistent, indicating that climate models capture many aspects of the externally forced climate change. Despite the overall agreement, several notable differences remain. Relative to ConvLSTM, climate models simulate stronger warming in the Northern Hemisphere, weaker intensification of subtropical sea-level pressure, and smaller declines in land precipitation, suggesting that current models may underestimate circulation-driven land drying under climate change.

Despite its strong performance, ConvLSTM has several important limitations. First, ConvLSTM is trained using climate model simulations. Although a suite of ten ConvLSTM models is developed to minimize the sensitivity to potential biases in individual models, systematic biases shared among models—particularly in their representation of internal

variability—could influence the ability of ConvLSTM to disentangle the forced and internal components of observations. Second, because ConvLSTM combines convolutional and LSTM networks, it remains challenging to fully understand the sources of its skill. Further studies are therefore needed to better understand the information learned by the network and the sources of its skill. Third, the training and evaluation of ConvLSTM assume that the ensemble mean provides an accurate estimate of the true forced response. Although this assumption is theoretically well justified, finite ensemble sizes inevitably introduce sampling errors into the estimated forced response and internal variability, which may in turn influence the training and evaluation of ConvLSTM. Despite these limitations, our results demonstrate that advances in machine learning can substantially improve the estimation of the forced climate response, leading to more robust attribution of historical climate change and a more rigorous evaluation of climate models.

**Acknowledgements**

This study was supported by Office of Science, U.S. Department of Energy Biological and Environmental Research as part of the Regional and Global Model Analysis program area. The Pacific Northwest National Laboratory (PNNL) is operated for DOE by Battelle Memorial Institute under contract DE-AC05-76RL01830. We acknowledge the WCRP Working Group on Coupled Modeling, which is responsible for CMIP, and the climate modeling groups for producing and making available their model outputs. This research used resources of the National Energy Research Scientific Computing Center, which is supported by the Office of Science of the U.S. Department of Energy under Contract No. DE-AC02-05CH1123. We also acknowledge the data access and computing support provided by the NCAR CMIP Analysis Platform (doi:10.5065/D60R9MSP).

**Completing interests**

The authors declare no competing interests.

**Data availability**

The CMIP6 outputs are available from the Earth System Grid Federation (ESGF) Portal at https://esgf-node.llnl.gov/search/cmip6/. The ERA5 reanalysis data is available from the Copernicus data store at cds.climate.copernicus.eu.

**Code availability**

The code of ConvLSTM and the python scripts to perform analyses and plot figures will be available upon publication.

**Author contributions**

W.Z. conceived the study. P.K. developed the ConvLSTM code. W.Z. and P.K. conducted the analysis. R.L., S-P.X., H.W. and F.Q. contributed to improving the analyses and interpretation. W.Z. drafted the manuscript and all authors edited the paper.

## Methods

### Data, preprocessing, training and evaluation

We trained and evaluated the ConvLSTM model using a large ensemble of climate model data (Table S1). The large-ensemble simulation data allows the training of ConvLSTM as the forced response is known from the ensemble mean. The data combines historical runs (pre-2015) with SSP scenario runs (post-2015). The size of ensemble runs and the SSP scenarios used by the models are shown in Table S1. The multi-model data captures the breadth of the available data from climate models with their individual representation of natural variability and forced response and therefore provides an excellent basis for training and evaluating ConvLSTM.

The three variables used for training and validation are surface temperature (Ts), sea-level pressure (SLP), and precipitation (Pr), all at a $2.5° \times 2.5°$ resolution. Prior to training, each realization $i$ is standardized at every grid point by subtracting its climatological mean ($\mu_i^{clim}$) and then dividing by $\sigma_i$, the standard deviation computed along the time axis. The three standardized variables are then stacked together and provided as input to ConvLSTM.

Skill evaluation: We trained a ConvLSTM model on all realizations from the first ten models in Table S1, using yearly data from 1950 to 2029. For early stopping, we monitored the validation error on a subset of 60 realizations. This ensures an optimal bias-variance tradeoff, preventing the model from overfitting to biases in the training data and thus improving its generalization to unseen models. The performance of the trained ConvLSTM is then evaluated using the five evaluation models (bottom five in Table S1):

1. ForceSMIP evaluation dataset, which includes one realization from each evaluation model. Notably, this dataset uses 2m temperature (Tas) while our ConvLSTM is trained on surface temperature (Ts).
2. Multiple realizations from the five evaluation models, considering the limited case size of the ForceSMIP evaluation dataset.

Observational inference: We construct 10 different ConvLSTM models to extract the forced response from our observational dataset. Each was trained on a distinct subset of 10 randomly selected models in Table S1. To ensure consistency with the observational data, the cases used to train and validate the 10 ConvLSTM models are standardized using the 1940-2024 climatology.

For observational inference, we use the historical evolution of Ts, SLP, and Pr from 1940 to 2024 obtained from ERA5 reanalysis data. Before being fed into ConvLSTM, these observational data are also standardized at each grid point by subtracting their climatology and dividing by their temporal standard deviation. The final extracted response is then de-standardized by reversing this process.

**Other Methods for Comparison with ConvLSTM**

Two baseline methods that are traditionally used to extract the forced response:

1. MMM (multi-model mean): The forced response is estimated as the mean of the forced response (ensemble run mean) simulated by each model in the training dataset.
2. EOF-1: The forced response is estimated as the leading global EOF.

Four of the representative methods from ForceSMIP:

3. LFCA (Wills et al. 2020): The forced response is estimated as the slowest evolving

spatial patterns extracted using linear discriminant analysis applied to the principal components.

4. TrainingEM (a scaled MMM method): The forced response is estimated based on the mean of the training models but rescaled so that it has the same GMST trend over 1950–2022 as the realization of interest.
5. 3D-UNet Fingerprints (Bône et al. 2024): A machine learning approach that utilizes a UNet to extract the forced response from a given realization.
6. EOF-SLR (Gavrilov et al. 2024): The forced response is estimated using a linear combination of optimal fingerprint patterns, which is based on an ensemble EOF decomposition on the training model data.

**ConvLSTM Architecture**

The left panel of Figure 1 illustrates the ConvLSTM's sequential data processing. At each time step t, the model receives anomaly fields of Ts, SLP, and Pr ($x_t$) as input and outputs the corresponding internal variability fields ($x_t^{iv}$). In addition to the input data, the network receives a hidden state ($h_{t-1}^i$) and a cell state ($c_{t-1}^i$) from the previous time step. The hidden state carries features from the preceding time step, while the cell state (or 'context') maintains a longer-term record of relevant past features. The specific model employed here consists of two stacked ConvLSTM cells, each with its own hidden and cell states. The final output is produced by passing the hidden state of the second cell ($h_t^2$) through a single 2D convolutional layer. The right panel of Figure 1 shows the internal structure of an individual ConvLSTM cell. First, the input data and the previous hidden state ($h_{t-1}^i$) are concatenated and processed by a convolutional layer to extract spatial features. These features are then managed by three gates–the forget and input gates use these features to update the cell's long-term memory, or context ($c_t^i$); the output gate combines the updated context and the previous hidden state to generate the new hidden state for the current time step ($h_t^i$). This flexible context maintenance allows the model to learn to retain pertinent information at varying lags. This capability makes the ConvLSTM architecture eminently appropriate for diagnosing internal variability, which involves modes that operate on diverse temporal and spatial scales.

## Hyperparameter Optimization

Hyperparameter optimization was performed using Optuna (Akiba et al., 2019), an open-source Python package that employs Bayesian optimization to find the optimum configuration of ConvLSTM, with the resulting architectural choices summarized in Table S2. The size and dilation of the convolutional kernel control the spatial resolution and receptive field of the features extracted by the convolutional operators, while the number of ConvLSTM layers sets the depth of the network and its capacity to learn hierarchical spatiotemporal representations of natural variability. The hidden size reflects the capacity of the hidden state ($h_t^i$) and the context ($c_t^i$) and represents a compromise between the richness of the memory representation and the limited availability of data. The learning rate and batch size are strongly coupled and represent a trade-off between training speed, memory usage, and convergence stability. Finally, dropout mitigates overfitting by introducing redundancy in the feature representations.

## Correction to better capture the transient forced response to volcanic eruptions

The original estimate of the forced response by ConvLSTM shows pronounced peaks in RMSE coinciding with the three major historical volcanic eruptions, indicating the difficulty in distinguishing the short-lived volcanic response from internal variability (Fig. S5). This deficiency is also seen in other methods and expected as volcanic forcing produces a transient climate response that interacts strongly with internal modes. Here, we remedy such a deficiency by identifying the robust biases of ConvLSTM during training and correcting them in the estimate of ConvLSTM. The detailed steps are as follows. First, the bias is computed for each training model at each year and grid point. Then, for each training model, we compute the robust bias as the mean of the bias that presents the same sign across over 80% of its realizations. Next, the final robust bias is computed as the mean of the individual-model bias that presents the same sign across 8 out of 10 training models. Finally, the mean robust bias is used to correct the original estimate of ConvLSTM. The correction is most effective over the periods of major volcanic eruptions, leading to a reduced RMSE over those periods (e.g., near 1963 for Agung and 1991 for Pinatubo) (Fig. S5). This correction only slightly affects the time-mean RMSE and the RMSE of the 1980-2022 trend, and ConvLSTM still presents better overall skills than other methods without the correction (Fig. S6).

### Metrics

Equatorial Pacific zonal temperature gradient: Defined as the difference in surface temperature between the west equatorial Pacific (7.5°S-7.5°N, 140°-210°) and the east equatorial Pacific (7.5°S-7.5°N, 210°-280°).

Atlantic temperature asymmetry: Defined as the difference in sea surface temperature between the north Atlantic (0-60°N, 290°-12.5°) and the south Atlantic (30°S-0, 290°-12.5°).

Northeast Pacific SLP: Defined as the mean sea-level pressure over the ocean within 25°N-52°N, 187.5°-235°.

Southwest North American Pr: Defined as the mean precipitation over the land within 27.5°N-45°N, 187.5°-255°.

The global or regional mean is computed with area weighting.

### Identification of IPO, AMV, and AM

With the forced and internal components estimated by ConvLSTM, internal modes can be cleanly identified from the internal components instead of from the full observations. Specifically, the modes of IPO, AMV, and AM are extracted as the leading regional EOF of the internal components of the Ts variability. IPO is identified as the leading EOF over the Pacific Ocean, 115°E-75°W and 60°S-60°N. AMV is identified as the leading EOF over the Atlantic Ocean, Atlantic, 70°W-20°E, 30°S-60°N. AM is identified as the leading EOF over the Arctic region north of 65°N.

### References:

Baek, Seung H., Yochanan Kushnir, Mingfang Ting, Jason E. Smerdon, and Juan M. Lora. 2022. "Regional Signatures of Forced North Atlantic SST Variability: A Limited Role for Aerosols and Greenhouse Gases." *Geophysical Research Letters* 49(8):e2022GL097794. doi:10.1029/2022GL097794.

Bône, Constantin, Guillaume Gastineau, Sylvie Thiria, Patrick Gallinari, and Carlos Mejia. 2024. "Separation of Internal and Forced Variability of Climate Using a U-Net."

*Journal of Advances in Modeling Earth Systems* 16(6):e2023MS003964. doi:10.1029/2023MS003964.

Corti, S., F. Molteni, and T. N. Palmer. 1999. “Signature of Recent Climate Change in Frequencies of Natural Atmospheric Circulation Regimes.” *Nature* 398:799–802. doi:10.1038/19745.

Dai, Aiguo, John C. Fyfe, Shang-Ping Xie, and Xingang Dai. 2015. “Decadal Modulation of Global Surface Temperature by Internal Climate Variability.” *Nature Climate Change* 5(6):555–59. doi:10.1038/nclimate2605.

Deser, C., F. Lehner, K. B. Rodgers, T. Ault, T. L. Delworth, P. N. DiNezio, A. Fiore, C. Frankignoul, J. C. Fyfe, D. E. Horton, J. E. Kay, R. Knutti, N. S. Lovenduski, J. Marotzke, K. A. McKinnon, S. Minobe, J. Randerson, J. A. Screen, I. R. Simpson, and M. Ting. 2020. “Insights from Earth System Model Initial-Condition Large Ensembles and Future Prospects.” *Nature Climate Change* 10(4):277–86. doi:10.1038/s41558-020-0731-2.

Deser, Clara, Michael A. Alexander, Shang-Ping Xie, and Adam S. Phillips. 2010. “Sea Surface Temperature Variability: Patterns and Mechanisms.” *Annual Review of Marine Science* 2(Volume 2, 2010):115–43. doi:10.1146/annurev-marine-120408-151453.

England, Matthew H., Shayne McGregor, Paul Spence, Gerald A. Meehl, Axel Timmermann, Wenju Cai, Alex Sen Gupta, Michael J. McPhaden, Ariaan Purich, and Agus Santoso. 2014. “Recent Intensification of Wind-Driven Circulation in the Pacific and the Ongoing Warming Hiatus.” *Nature Climate Change* 4(3):222–27. doi:10.1038/nclimate2106.

Eyring, Veronika, Sandrine Bony, Gerald A. Meehl, Catherine A. Senior, Bjorn Stevens, Ronald J. Stouffer, and Karl E. Taylor. 2016. “Overview of the Coupled Model Intercomparison Project Phase 6 (CMIP6) Experimental Design and Organization.” *Geoscientific Model Development* 9(5):1937–58. doi:10.5194/gmd-9-1937-2016.

Fyfe, John C., Nathan P. Gillett, and Francis W. Zwiers. 2013. “Overestimated Global Warming over the Past 20 Years.” *Nature Climate Change* 3(9):767–69. doi:10.1038/nclimate1972.

Gavrilov, Andrey, Sergey Kravtsov, Maria Buyanova, Dmitry Mukhin, Evgeny Loskutov, and Alexander Feigin. 2024. “Forced Response and Internal Variability in Ensembles of Climate Simulations: Identification and Analysis Using Linear Dynamical Mode Decomposition.” *Climate Dynamics* 62(3):1783–1810. doi:10.1007/s00382-023-06995-1.

Giannini, A., R. Saravanan, and P. Chang. 2003. “Oceanic Forcing of Sahel Rainfall on Interannual to Interdecadal Time Scales.” *Science* 302(5647):1027–30. doi:10.1126/science.1089357.

Goldenberg, Stanley B., Christopher W. Landsea, Alberto M. Mestas-Nuñez, and William M. Gray. 2001. “The Recent Increase in Atlantic Hurricane Activity: Causes and Implications.” *Science* 293(5529):474–79. doi:10.1126/science.1060040.

He, Chengfei, Amy C. Clement, Sydney M. Kramer, Mark A. Cane, Jeremy M. Klavans, Tyler M. Fenske, and Lisa N. Murphy. 2023. “Tropical Atlantic Multidecadal Variability Is Dominated by External Forcing.” *Nature* 622(7983):521–27. doi:10.1038/s41586-023-06489-4.

Hochreiter, Sepp, and Jürgen Schmidhuber. 1997. “Long Short-Term Memory.” *Neural Computation* 9(8):1735–80. doi:10.1162/neco.1997.9.8.1735.

Holland, M. M., and C. M. Bitz. 2003. “Polar Amplification of Climate Change in Coupled Models.” *Climate Dynamics* 21(3):221–32. doi:10.1007/s00382-003-0332-6.

Hu, Y., and Q. Fu. 2007. “Observed Poleward Expansion of the Hadley Circulation since 1979.” *Atmospheric Chemistry and Physics* 7(19):5229–36. doi:10.5194/acp-7-5229-2007.

Hua, Wenjian, and Aiguo Dai. 2024. “Impacts of External Forcing and Internal Variability on the North Atlantic Multidecadal Variations since 1950.” *Science China Earth Sciences* 67(8):2699–2701. doi:10.1007/s11430-024-1378-y.

Jacobson, Tess W. P., Richard Seager, A. Park Williams, Isla R. Simpson, Karen A. McKinnon, and Haibo Liu. 2024. “An Unexpected Decline in Spring Atmospheric Humidity in the Interior Southwestern United States and Implications for Forest Fires.” doi:10.1175/JHM-D-23-0121.1.

Klavans, Jeremy M., Pedro N. DiNezio, Amy C. Clement, Clara Deser, Timothy M. Shanahan, and Mark A. Cane. 2025. “Human Emissions Drive Recent Trends in North Pacific Climate Variations.” *Nature* 644(8077):684–92. doi:10.1038/s41586-025-09368-2.

Knight, Jeff R., Chris K. Folland, and Adam A. Scaife. 2006. “Climate Impacts of the Atlantic Multidecadal Oscillation.” *Geophysical Research Letters* 33(17). doi:10.1029/2006GL026242.

Kosaka, Yu, and Shang-Ping Xie. 2013. “Recent Global-Warming Hiatus Tied to Equatorial Pacific Surface Cooling.” *Nature* 501(7467):403–7. doi:10.1038/nature12534.

Lecun, Y., L. Bottou, Y. Bengio, and P. Haffner. 1998. “Gradient-Based Learning Applied to Document Recognition.” *Proceedings of the IEEE* 86(11):2278–2324. doi:10.1109/5.726791.

Manabe, Syukuro, and Ronald J. Stouffer. 1980. “Sensitivity of a Global Climate Model to an Increase of CO2 Concentration in the Atmosphere.” *Journal of Geophysical Research: Oceans* 85(C10):5529–54. doi:10.1029/JC085iC10p05529.

Masson-Delmotte, Valérie, Panmao Zhai, Anna Pirani, Sarah L. Connors, Clotilde Péan, Sophie Berger, Nada Caud, Yang Chen, Leah Goldfarb, Melissa I. Gomis, Mengtian Huang, Katherine Leitzell, Elisabeth Lonnoy, J. B. Robin Matthews, Thomas K. Maycock, Tim Waterfield, Özge Yelekçi, Rong Yu, and Botao Zhou, eds. 2021. *Climate Change 2021: The Physical Science Basis. Contribution of Working Group I to the Sixth Assessment Report of the Intergovernmental Panel on Climate Change*. Cambridge, United Kingdom and New York, NY, USA: Cambridge University Press.

Meehl, Gerald A., Julie M. Arblaster, John T. Fasullo, Aixue Hu, and Kevin E. Trenberth. 2011. “Model-Based Evidence of Deep-Ocean Heat Uptake during Surface-Temperature Hiatus Periods.” *Nature Climate Change* 1(7):360–64. doi:10.1038/nclimate1229.

Newman, Matthew, Michael A. Alexander, Toby R. Ault, Kim M. Cobb, Clara Deser, Emanuele Di Lorenzo, Nathan J. Mantua, Arthur J. Miller, Shoshiro Minobe, Hisashi Nakamura, Niklas Schneider, Daniel J. Vimont, Adam S. Phillips, James D. Scott, and Catherine A. Smith. 2016. “The Pacific Decadal Oscillation, Revisited.” *Journal of Climate* 29(12):4399–4427. doi:10.1175/JCLI-D-15-0508.1.

Otterå, Odd Helge, Mats Bentsen, Helge Drange, and Lingling Suo. 2010. “External Forcing as a Metronome for Atlantic Multidecadal Variability.” *Nature Geoscience* 3(10):688–94. doi:10.1038/ngeo955.

Philander, S. G. H. 1983. “El Niño Southern Oscillation Phenomena.” *Nature* 302(5906):295–301. doi:10.1038/302295a0.

Rantanen, Mika, Alexey Yu Karpechko, Antti Lipponen, Kalle Nordling, Otto Hyvärinen, Kimmo Ruosteenoja, Timo Vihma, and Ari Laaksonen. 2022. “The Arctic Has Warmed Nearly Four Times Faster than the Globe since 1979.” *Communications Earth & Environment* 3(1):1–10. doi:10.1038/s43247-022-00498-3.

Seager, Richard, Mark Cane, Naomi Henderson, Dong-Eun Lee, Ryan Abernathey, and Honghai Zhang. 2019. “Strengthening Tropical Pacific Zonal Sea Surface Temperature Gradient Consistent with Rising Greenhouse Gases.” *Nature Climate Change* 9(7):517–22. doi:10.1038/s41558-019-0505-x.

Serreze, M. C., A. P. Barrett, J. C. Stroeve, D. N. Kindig, and M. M. Holland. 2009. "The Emergence of Surface-Based Arctic Amplification." *The Cryosphere* 3(1):11–19. doi:10.5194/tc-3-11-2009.

Shi, Xingjian, Zhourong Chen, Hao Wang, Dit-Yan Yeung, Wai-kin Wong, and Wang-chun Woo. 2015. "Convolutional LSTM Network: A Machine Learning Approach for Precipitation Nowcasting."

Shindell, Drew T. 2014. "Inhomogeneous Forcing and Transient Climate Sensitivity." *Nature Climate Change* 4(4):274–77. doi:10.1038/nclimate2136.

Sweeney, Aodhan J., Qiang Fu, Stephen Po-Chedley, Hailong Wang, and Muyin Wang. 2023. "Internal Variability Increased Arctic Amplification During 1980–2022." *Geophysical Research Letters* 50(24):e2023GL106060. doi:10.1029/2023GL106060.

Sweeney, Aodhan J., Qiang Fu, Stephen Po-Chedley, Hailong Wang, and Muyin Wang. 2024. "Unique Temperature Trend Pattern Associated With Internally Driven Global Cooling and Arctic Warming During 1980–2022." *Geophysical Research Letters* 51(11):e2024GL108798. doi:10.1029/2024GL108798.

Timmermann, Axel, Soon-Il An, Jong-Seong Kug, Fei-Fei Jin, Wenju Cai, Antonietta Capotondi, Kim M. Cobb, Matthieu Lengaigne, Michael J. McPhaden, Malte F. Stuecker, Karl Stein, Andrew T. Wittenberg, Kyung-Sook Yun, Tobias Bayr, Han-Ching Chen, Yoshimitsu Chikamoto, Boris Dewitte, Dietmar Dommenget, Pamela Grothe, Eric Guilyardi, Yoo-Geun Ham, Michiya Hayashi, Sarah Ineson, Daehyun Kang, Sunyong Kim, WonMoo Kim, June-Yi Lee, Tim Li, Jing-Jia Luo, Shayne McGregor, Yann Planton, Scott Power, Harun Rashid, Hong-Li Ren, Agus Santoso, Ken Takahashi, Alexander Todd, Guomin Wang, Guojian Wang, Ruihuang Xie, Woo-Hyun Yang, Sang-Wook Yeh, Jinho Yoon, Elke Zeller, and Xuebin Zhang. 2018. "El Niño–Southern Oscillation Complexity." *Nature* 559(7715):535–45. doi:10.1038/s41586-018-0252-6.

Ting, Mingfang, Yochanan Kushnir, Richard Seager, and Cuihua Li. 2009. "Forced and Internal Twentieth-Century SST Trends in the North Atlantic." *Journal of Climate* 22(6):1469–81. doi:10.1175/2008JCLI2561.1.

Vecchi, Gabriel A., Brian J. Soden, Andrew T. Wittenberg, Isaac M. Held, Ants Leetmaa, and Matthew J. Harrison. 2006. "Weakening of Tropical Pacific Atmospheric Circulation Due to Anthropogenic Forcing." *Nature* 441(7089):73–76. doi:10.1038/nature04744.

Watanabe, Masahiro, Jean-Louis Dufresne, Yu Kosaka, Thorsten Mauritsen, and Hiroaki Tatebe. 2021. "Enhanced Warming Constrained by Past Trends in Equatorial Pacific Sea Surface Temperature Gradient." *Nature Climate Change* 11(1):33–37. doi:10.1038/s41558-020-00933-3.

Watanabe, Masahiro, Sarah M. Kang, Matthew Collins, Yen-Ting Hwang, Shayne McGregor, and Malte F. Stuecker. 2024. “Possible Shift in Controls of the Tropical Pacific Surface Warming Pattern.” *Nature* 630(8016):315–24. doi:10.1038/s41586-024-07452-7.

Watanabe, Masahiro, Hideo Shiogama, Hiroaki Tatebe, Michiya Hayashi, Masayoshi Ishii, and Masahide Kimoto. 2014. “Contribution of Natural Decadal Variability to Global Warming Acceleration and Hiatus.” *Nature Climate Change* 4(10):893–97. doi:10.1038/nclimate2355.

Wills, Robert C. J., David S. Battisti, Kyle C. Armour, Tapio Schneider, and Clara Deser. 2020. “Pattern Recognition Methods to Separate Forced Responses from Internal Variability in Climate Model Ensembles and Observations.” *Journal of Climate* 33(20):8693–8719. doi:10.1175/JCLI-D-19-0855.1.

Wills, Robert C. J., Clara Deser, Karen A. McKinnon, Adam Phillips, Stephen Po-Chedley, Sebastian Sippel, Anna L. Merrifield, Constantin Bône, Céline Bonfils, Gustau Camps-Valls, Stephen Cropper, Charlotte Connolly, Shiheng Duan, Homer Durand, Alexander Feigin, M. A. Fernandez, Guillaume Gastineau, Andrei Gavrilov, Emily Gordon, Moritz Günther, Maren Höver, Sergey Kravtsov, Yan-Ning Kuo, Justin Lien, Gavin D. Madakumbura, Nathan Mankovich, Matthew Newman, Jamin Rader, Jia-Rui Shi, Sang-Ik Shin, and Gherardo Varando. 2026. “Forced Component Estimation Statistical Method Intercomparison Project (ForceSMIP).” *Journal of Climate* 39(8):1927–53. doi:10.1175/JCLI-D-25-0326.1.

Wills, Robert C. J., Yue Dong, Cristian Proistosecu, Kyle C. Armour, and David S. Battisti. 2022. “Systematic Climate Model Biases in the Large-Scale Patterns of Recent Sea-Surface Temperature and Sea-Level Pressure Change.” *Geophysical Research Letters* 49(17):e2022GL100011. doi:10.1029/2022GL100011.

Xie, Shang-Ping, and James A. Carton. 2004. “Tropical Atlantic Variability: Patterns, Mechanisms, and Impacts.” Pp. 121–42 in *Earth’s Climate*. American Geophysical Union (AGU).

Xie, Shang-Ping, Clara Deser, Gabriel A. Vecchi, Matthew Collins, Thomas L. Delworth, Alex Hall, Ed Hawkins, Nathaniel C. Johnson, Christophe Cassou, Alessandra Giannini, and Masahiro Watanabe. 2015. “Towards Predictive Understanding of Regional Climate Change.” *Nature Climate Change* 5(10):921–30. doi:10.1038/nclimate2689.

Xie, Shang-Ping, A. Miyamoto, Qihua Peng, and M. T. Luongo. 2026. “Major Role for Internal Variability in Tropical Pacific Warming Pattern over Satellite Era.” 123.

Zhang, Rong, and Thomas L. Delworth. 2006. “Impact of Atlantic Multidecadal Oscillations on India/Sahel Rainfall and Atlantic Hurricanes.” *Geophysical Research Letters* 33(17). doi:10.1029/2006GL026267.

Zhang, Rong, Rowan Sutton, Gokhan Danabasoglu, Young-Oh Kwon, Robert Marsh, Stephen G. Yeager, Daniel E. Amrhein, and Christopher M. Little. 2019. “A Review of the Role of the Atlantic Meridional Overturning Circulation in Atlantic Multidecadal Variability and Associated Climate Impacts.” *Reviews of Geophysics* 57(2):316–75. doi:10.1029/2019RG000644.

Zhou, Wenyu, L. Ruby Leung, Chuan-Chieh Chang, Ming Zhao, Huang-Hsiung Hsu, Hsin-Chien Liang, Chia-Ying Tu, Karthik Balaguru, and Jian Lu. 2026. “Poleward Migration of Tropical Cyclones over 1980–2024 Is Dominated by Pacific Variability.” *Nature Geoscience* 19(1):42–51. doi:10.1038/s41561-025-01866-2.

Zhou, Wenyu, L. Ruby Leung, Bryce E. Harrop, Ziming Chen, and Chuan-Chieh Chang. 2026. “A Theoretical Index for Understanding Distinct Land Relative Humidity Trends in Observations, Reanalyses, and Models.” *Proceedings of the National Academy of Sciences* 123(10):e2512645123. doi:10.1073/pnas.2512645123.

Zhou, Wenyu, L. Ruby Leung, and Jian Lu. 2024. “Steady Threefold Arctic Amplification of Externally Forced Warming Masked by Natural Variability.” *Nature Geoscience* 1–8. doi:10.1038/s41561-024-01441-1.

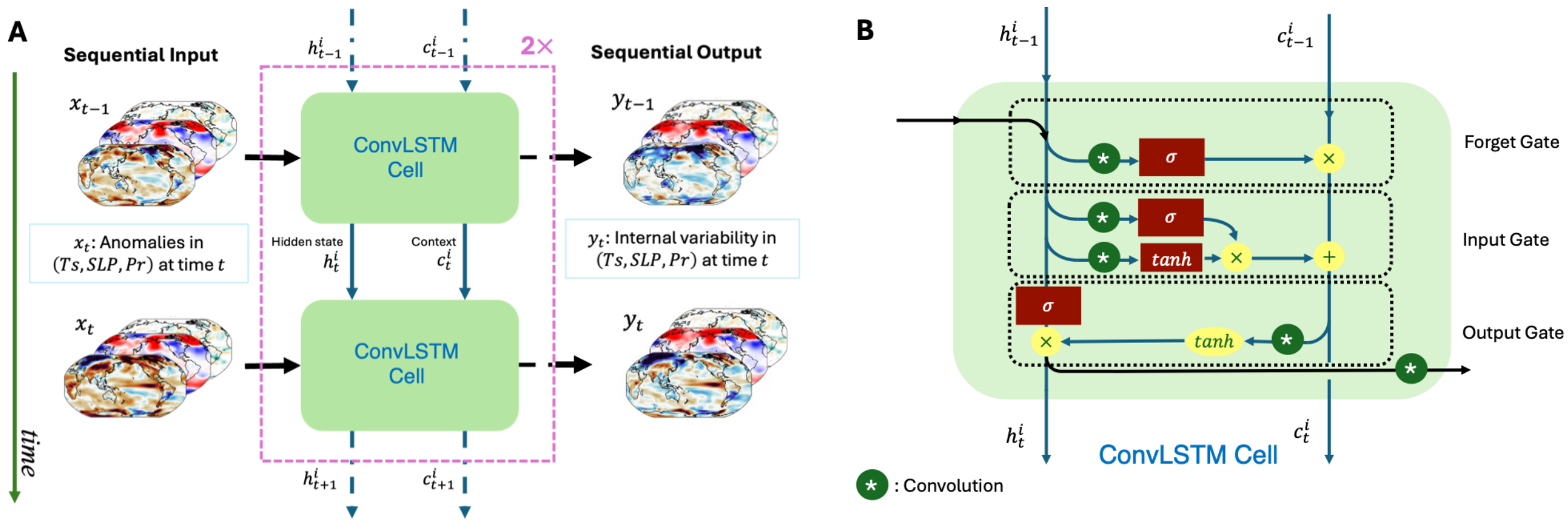

**Figure 1: Architecture of ConvLSTM.** (A) Sequential pipeline of ConvLSTM through time with inputs of the full anomaly of (*Ts, SLP, Pr*) and outputs of the internal component. (B) Gate structure that updates the context and hidden state within the ConvLSTM cell. The hidden state ($h_t^i$) and context ($c_t^i$) streams retain 'memory' of the past and inform ConvLSTM.

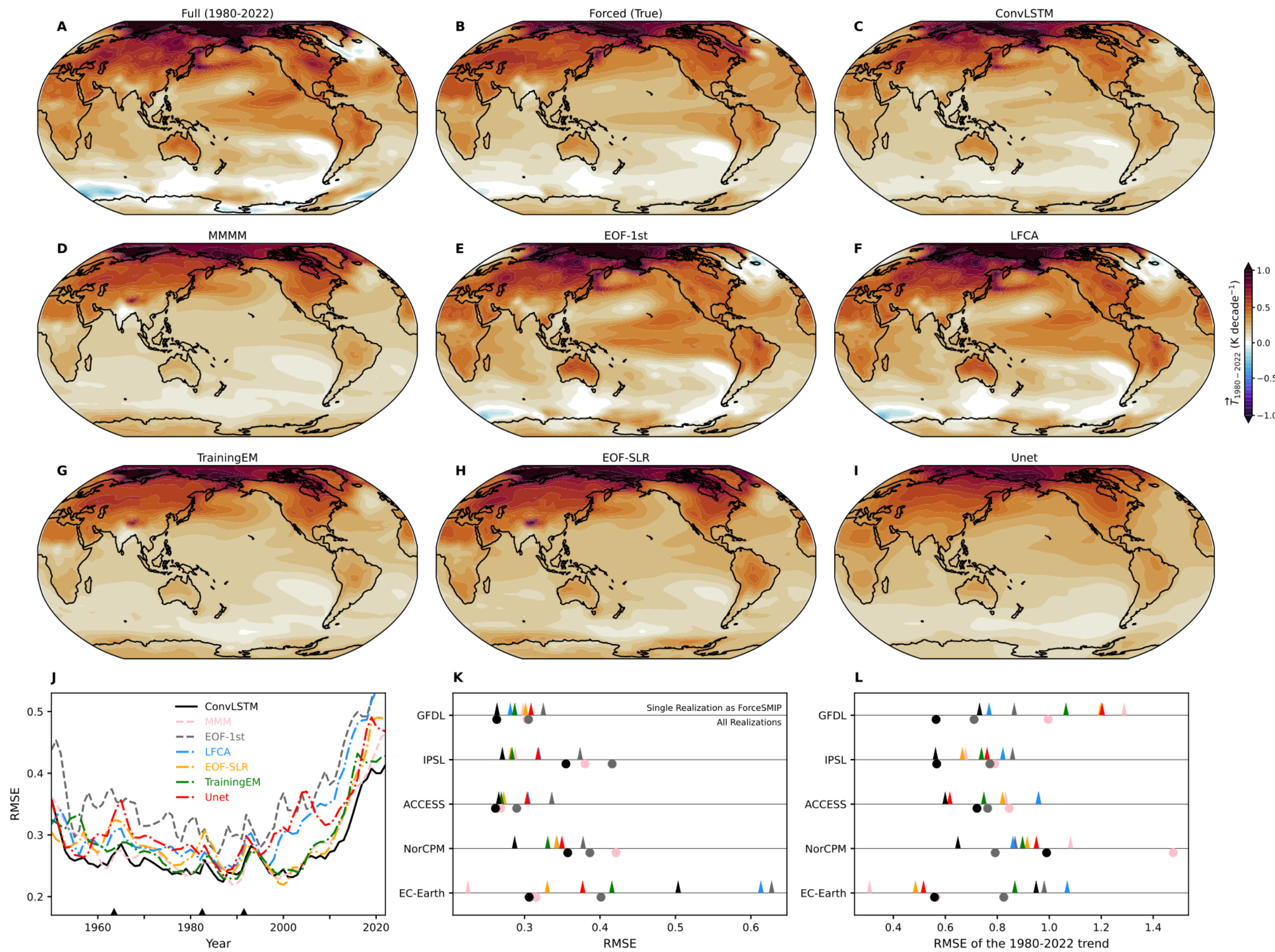

**Figure 2: Evaluation of different methods in capturing the forced response of surface temperature based on large-ensemble simulations of climate models.** (A) The full temperature trend over 1980-2022 in the evaluation case 1B (ACCESS-ESM1-5 r1). (B) The true forced trend, inferred from the 40-member ensemble mean of ACCESS-ESM1-5. (C-I) The forced trend estimated by ConvLSTM, MMM, EOF-1st, LFCA, TrainingEM, EOF-SLR, and UNet. (J) Time series of the global-mean (area weighted) RMSE of the estimated forced response relative to the true forced response, shown as the median across the five evaluation cases for each method. (K) Time-mean global-mean RMSE of the forced response estimated by each method (triangles for using only one realization of each evaluation model, as in ForceSMIP, and dots for using multiple realizations). (L) Same as K but for the global-mean RMSE of the forced trend over 1980-2022. The colors in K and L share the same legends as in J. To inform how good the estimate is relative to the raw estimate directly using the full anomaly, we have normalized RMSE of each method by RMSE of the raw estimate so that RMSE is unitless. The up-pointing triangle marks in J denote the time of three major volcanic eruptions.

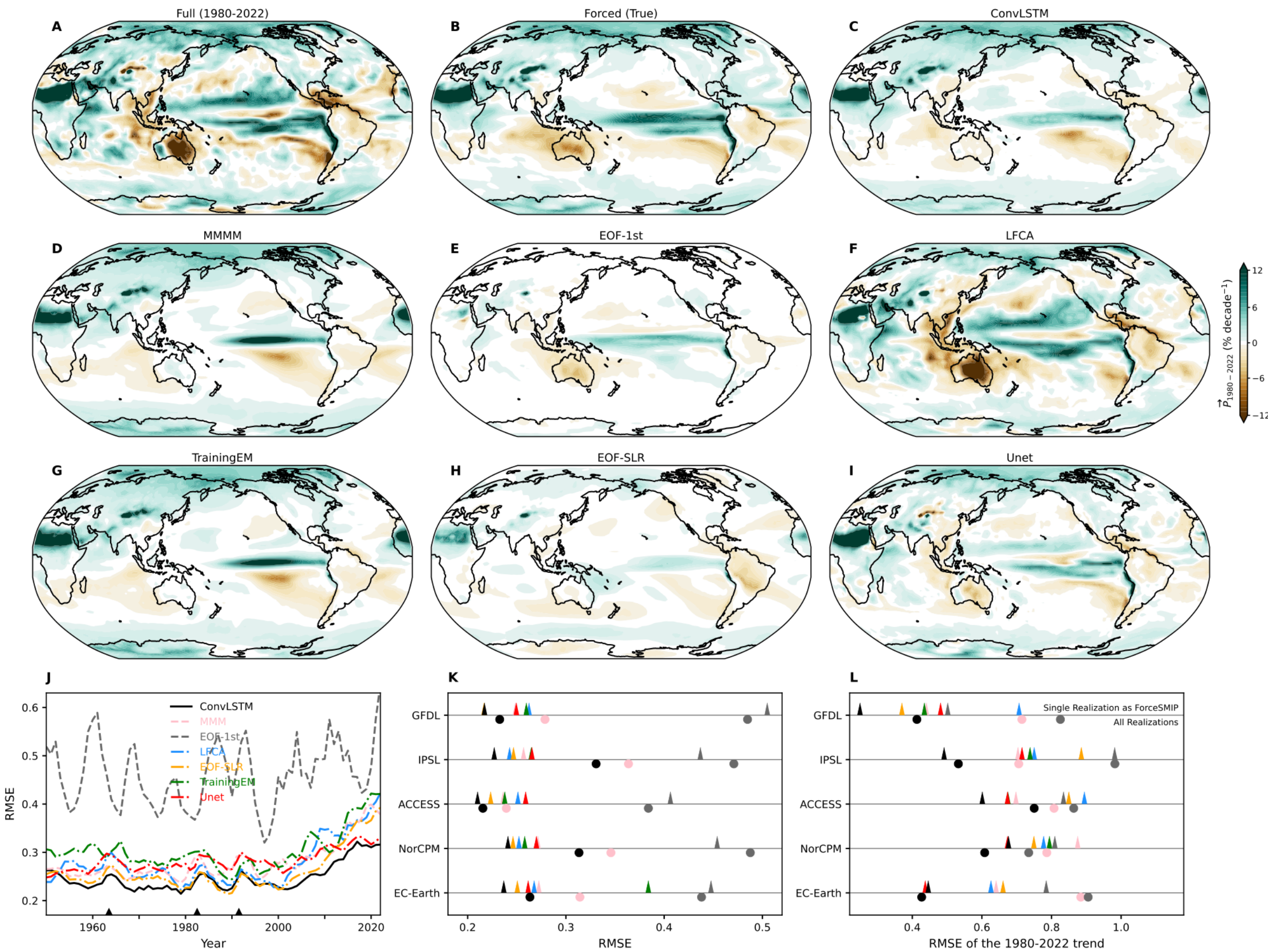

**Figure 3:** Same as Fig. 2 but for precipitation.

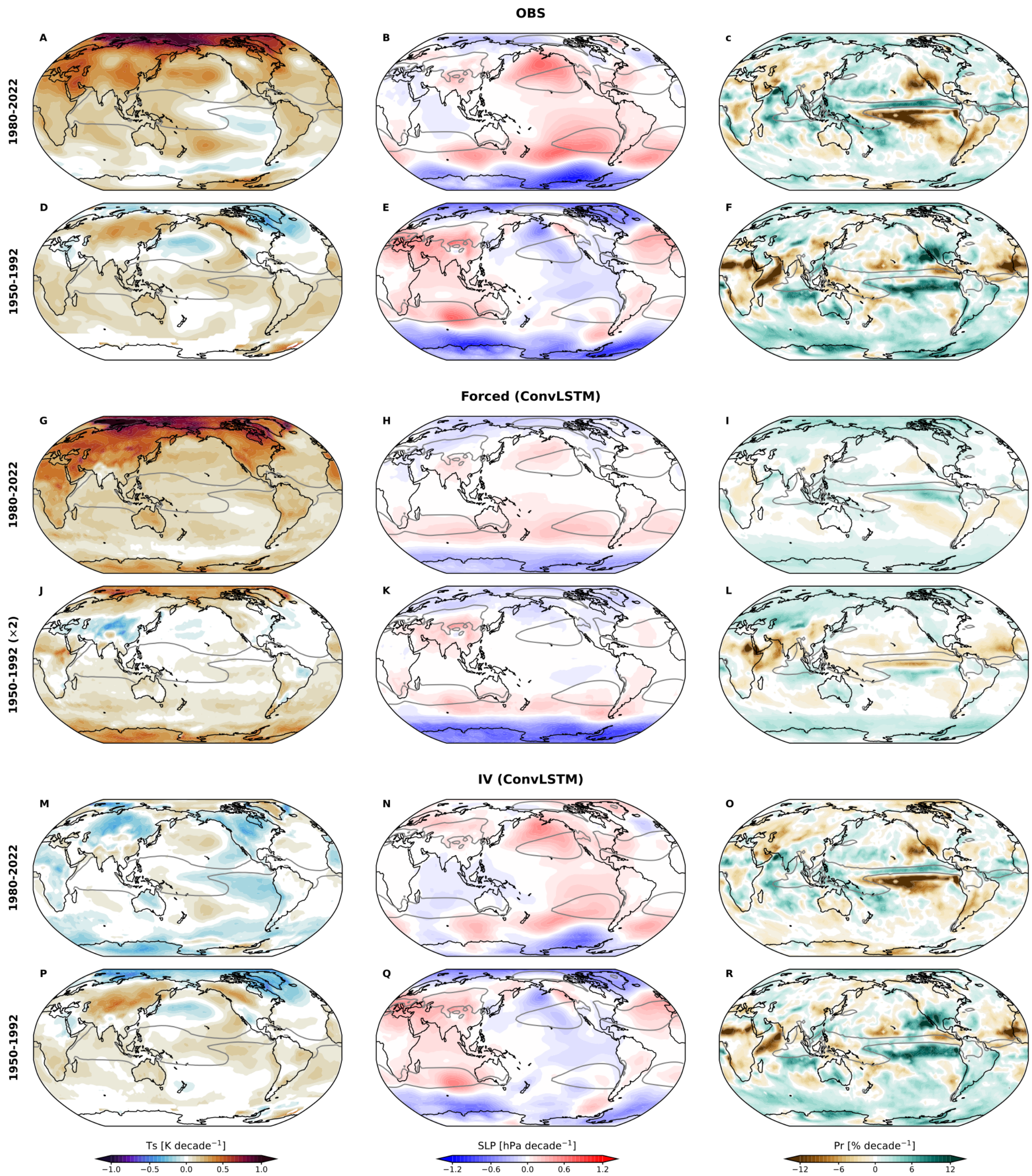

**Figure 4: Attribution of the observed trends to the forced and internal components.** (A-F) The observed trends (OBS) in surface temperature (Ts), sea-level pressure (SLP), and precipitation (Pr) over 1980-2022 and 1950-1992, respectively. (G-L) Same as A-F but for the forced trends estimated by ConvLSTM. (M-R) Same as A-F but for internal variability measured as IV = OBS – Forced. The black contours denote the climatology of Ts at 27$^{o}$C, SLP at 1016.5 hPa, and Pr at 6 mm day$^{-1}$.

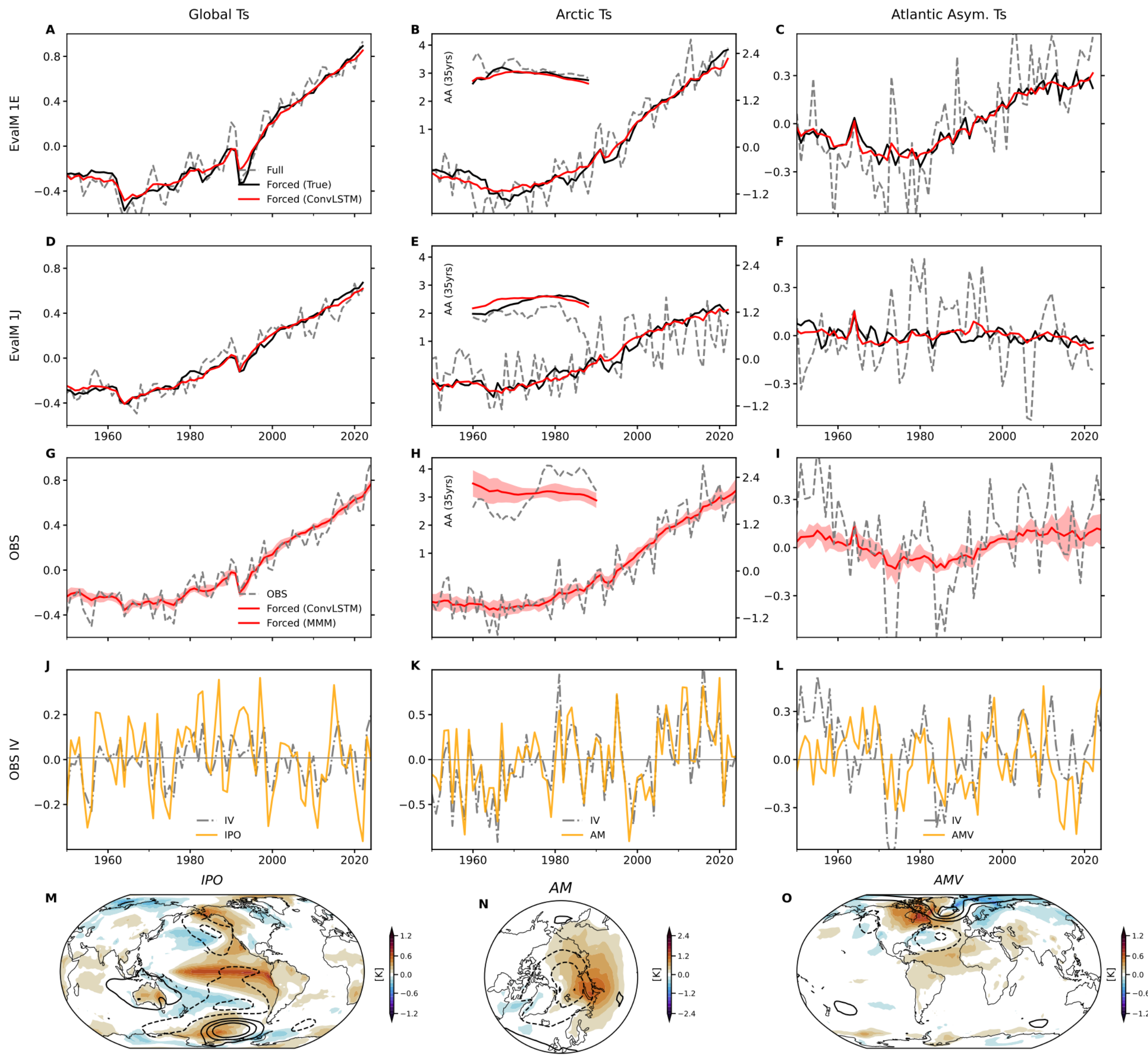

**Figure 5: Contributions of external forcing and internal variability to historical evolutions in global and regional temperature.** (A) Time series of the global mean temperature (anomaly relative to the 1960-2020 climatology) (dashed), the true forced component (black), and the ConvLSTM-derived forced component (red) in the evaluation case 1E (GFDL-SPEAR r3). (B) Same as (A) but for the Arctic (north of 65°N) mean temperature. The degree of Arctic Amplification (AA, measured by the warming ratio between the Arctic and global mean) is shown for moving 35-year periods (starting year indicated by x-axis). (C) Same as (A) but for the Atlantic inter-hemispheric temperature asymmetry (0-60°N minus 0-30°S). (D-F) Same as A-C but for the evaluation case 1J (NorCPM1 r4). (G-I) Same as A-C but for the observed time series (dashed) and the forced response estimated by ConvLSTM (red). The red shading indicates the root-sum-square of the standard deviation among the ten ConvLSTM models and the error of ConvLSTM

in the evaluation model datasets. (J) Internal variability (IV) of the global mean temperature evolution in observations estimated by ConvLSTM and PC of the interdecadal Pacific oscillation (IPO, derived as the first EOF of IV over the Pacific, 115$^{o}$E-75$^{o}$W and 60$^{o}$S-60$^{o}$N). (K) Internal variability of the Arctic mean temperature evolution and PC of the Arctic mode (AM, derived as the first EOF of IV north of 65$^{o}$N). (L) Internal variability of the Atlantic temperature asymmetry and PC of the Atlantic multidecadal variability (AMV, derived as the first EOF of IV over the Atlantic, 70$^{o}$W-20$^{o}$E, 30$^{o}$S-60$^{o}$N). (M-O) Spatial patterns of surface temperature (shading) and sea-level pressure (contours at 0.4hPa) for IPO, AM, and AMV. The unit is K for temperature, whereas AA and PCs are unitless.

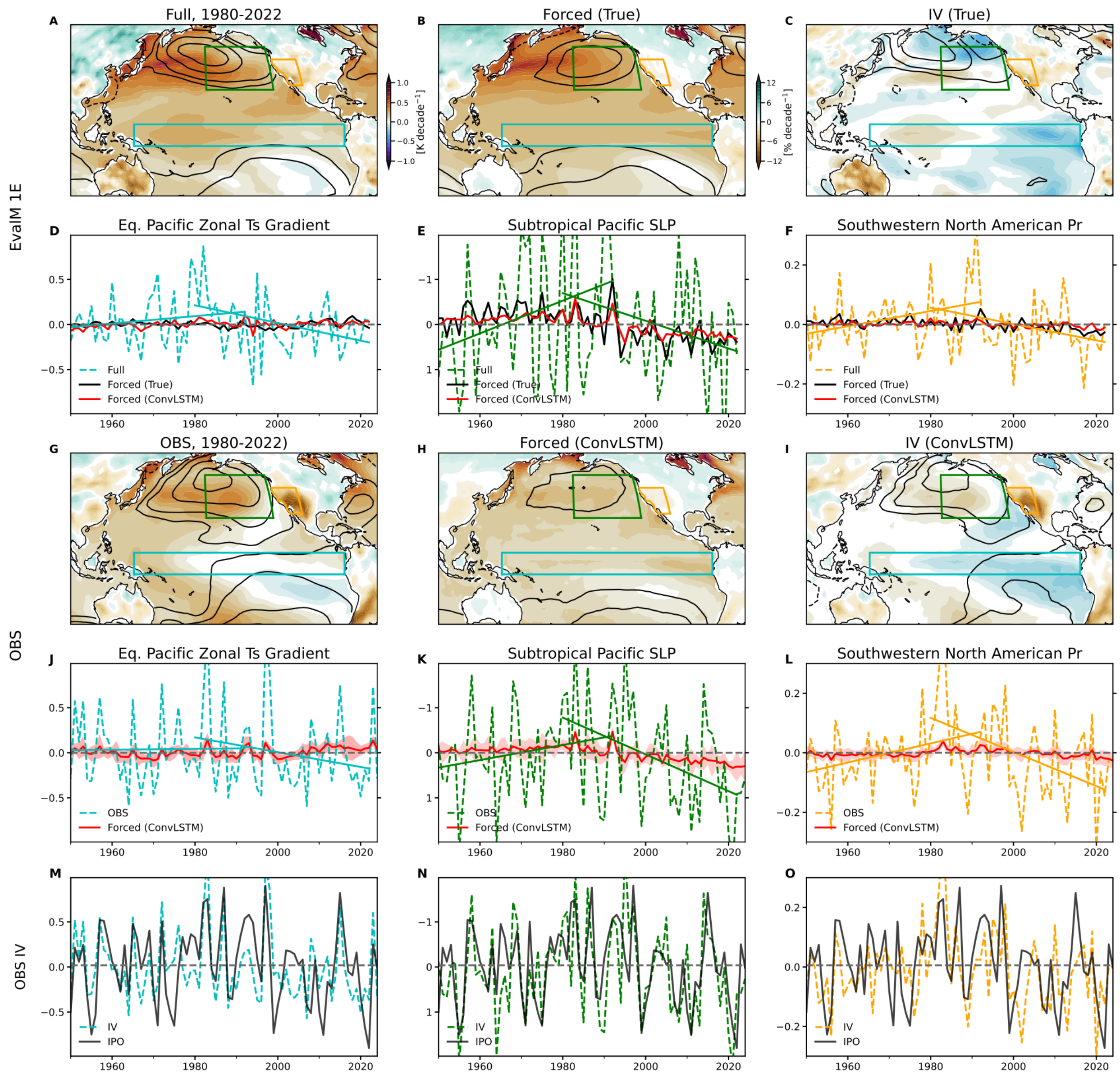

**Figure 6: Forced and internal components of historical variability around the Pacific.** (A-C) The full trends, the true forced trends, and the trends contributed by internal variability in Ts (shading over ocean), SLP (contours at 0.1hPa dec$^{-1}$), and Pr (shading over land) over 1980-2022 in the evaluation case 1E. The regions of the equatorial Pacific, the northeast subtropical Pacific, and the southwestern North American are denoted by colored boxes. (D) Time series of the zonal surface temperature gradient over equatorial Pacific (dashed), the true forced component (black), and the forced component estimated by ConvLSTM (red). The contrasting trends over 1950-1992 and 1980-2022 are denoted. (E-F) Same as D but for the northeast subtropical Pacific SLP and southwestern North American Pr. (G-I) Same as A-C but for the observed trends, the forced trends estimated by ConvLSTM, and the trends attributed to internal variability. (J-L) Same as D-F but for the observed time series (dashed) and the forced response estimated by ConvLSTM (red). The red shading indicates the root-sum-square of the standard deviation among the ten ConvLSTM models and the error of ConvLSTM in the evaluation model datasets. (M-O) Historical time series of IPO and internal variability of the equatorial Pacific Ts gradient, the northeast subtropical Pacific SLP, and the southwestern North American Pr.

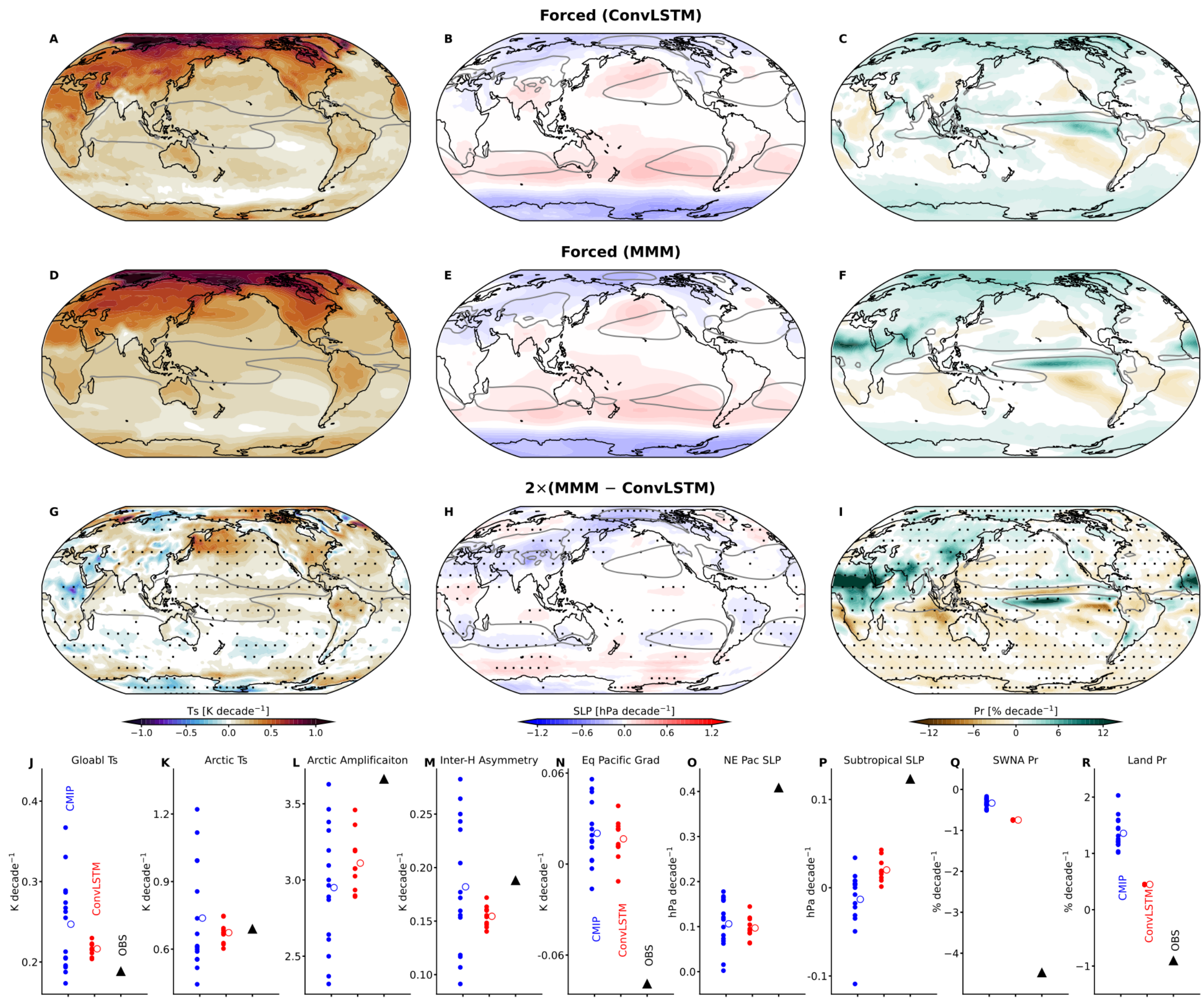

**Figure 7: Consistency and difference between the forced trends estimated by ConvLSTM and climate models.** (A-C) The forced trends over 1980-2022 in Ts, SLP, and Pr estimated by ConvLSTM. (D-F) Same as A-C but for the forced trends simulated by MMM of climate models, (G-I) The differences in the model-simulated and ConvLSTM-estimated forced trends. (J-R) Comparisons of the forced trends simulated by climate models (15 models; blue), the forced trends estimated by ConvLSTM (ten members; red), and the observed trends (OBS; black diamond) for the global warming rate, the Arctic warming rate, the degree of Arctic Amplification, the interhemispheric temperature asymmetry, the equatorial Pacific zonal Ts gradient, the northeast Pacific SLP, the subtropical zonal SLP, the southwestern North American Pr, and the global land mean Pr (Methods). The open circles represent the ensemble means.

**Supplementary Information for**

# Disentangling forced and internal components in observed climate evolution using spatiotemporal deep learning

Parvathi Kooloth[1], Wenyu Zhou[1*], L. Ruby Leung[1],
Shang-Ping Xie[2], Hailong Wang[1], and Qiang Fu[3]
[1]Pacific Northwest National Laboratory, Richland, WA, USA 99352
[2]University of California, San Diego, La Jolla, CA, USA 92093
[3]University of Washington, Seattle, WA, USA 98195
Corresponding Author: wenyu.zhou@pnnl.gov

This file contains:

Table S1-S2

Figure S1-S6

| Model | Members | Future scenario |
|---|---|---|
| CanESM5 | 25 | SSP370 |
| CESM2 | 50 | SSP370 |
| MIROC-ES2L | 10 | SSP370 |
| MIROC6 | 10 | SSP370 |
| MPI-ESM1-2-LR | 50 | SSP370 |
| CanESM5-1 | 10 | SSP370 |
| CESM2smbb | 50 | SSP370 |
| E3SM-2-0 | 21 | SSP370 |
| GISS-E2-1-G | 21 | SSP370 |
| MPI-ESM1-2-HR | 10 | SSP370 |
| ACCESS-ESM1-5 | 50 | SSP370 |
| EC-Earth3 | 23 | SSP245 |
| GFDL-SPREAR | 30 | SSP585 |
| IPSL-CM6A-LR | 11 | SSP370 |
| NorCPM1 | 15 | SSP245 |

| Five evaluation cases used in ForceSMIP | |
|---|---|
| ACCESS-ESM1-5 | 1B (r10i1p1f1) |
| EC-Earth3 | 1D (r6i1p1f1) |
| GFDL-SPEAR | 1E (r3i1p1f1) |
| IPSL-CM6A-LR | 1G (r3i1p1f1) |
| NorCPM1 | 1J (r4i1p1f1) |

**Table S1:** Large-ensemble simulations of multiple climate models used to train and evaluate the Convolutional LSTM. When compared with methods in ForceSMIP, ConvLSTM is trained using 10 models (top 10) and evaluated using the same single realization of the five evaluation models (bottom five) as ForceSMIP (see their Table 2). Considering the limited size of evaluation cases, we further evaluate the skills of ConvLSTM, MMM and EOF-1st based on multiple realizations of the five evaluation models. When applied to observations, ten ConvLSTM models are used, each trained on a distinct subset of ten climate models from the 15 models.

| Feature | Optimal Parameter |
|---|---|
| Learning rate | ~0.005 |
| Batch size | 8 |
| ConvLSTM Layers | 2 |
| Hidden size | 48 |
| Kernel size | 5 |
| Dilation | 3 |
| Dropout | 0.5 |

Table S2: Optimal Parameters for ConvLSTM

**Outlier behavior of EC-Earth3 r6**

EC-Earth3 r6 represents an extreme outlier of historical simulation, exhibiting a global warming trend nearly twice of the forced trend and an outlier Atlantic Meridional Overturning Circulation (AMOC) variability. We therefore quesiton whether r6 represents a realistic analogue of observed climate variability and whether it is an appropriate benchmark for method evaluation. In Fig. S2 of the ForceSMIP paper, one can also see that methods performing best for EC-Earth3 often perform among the worst for other models.

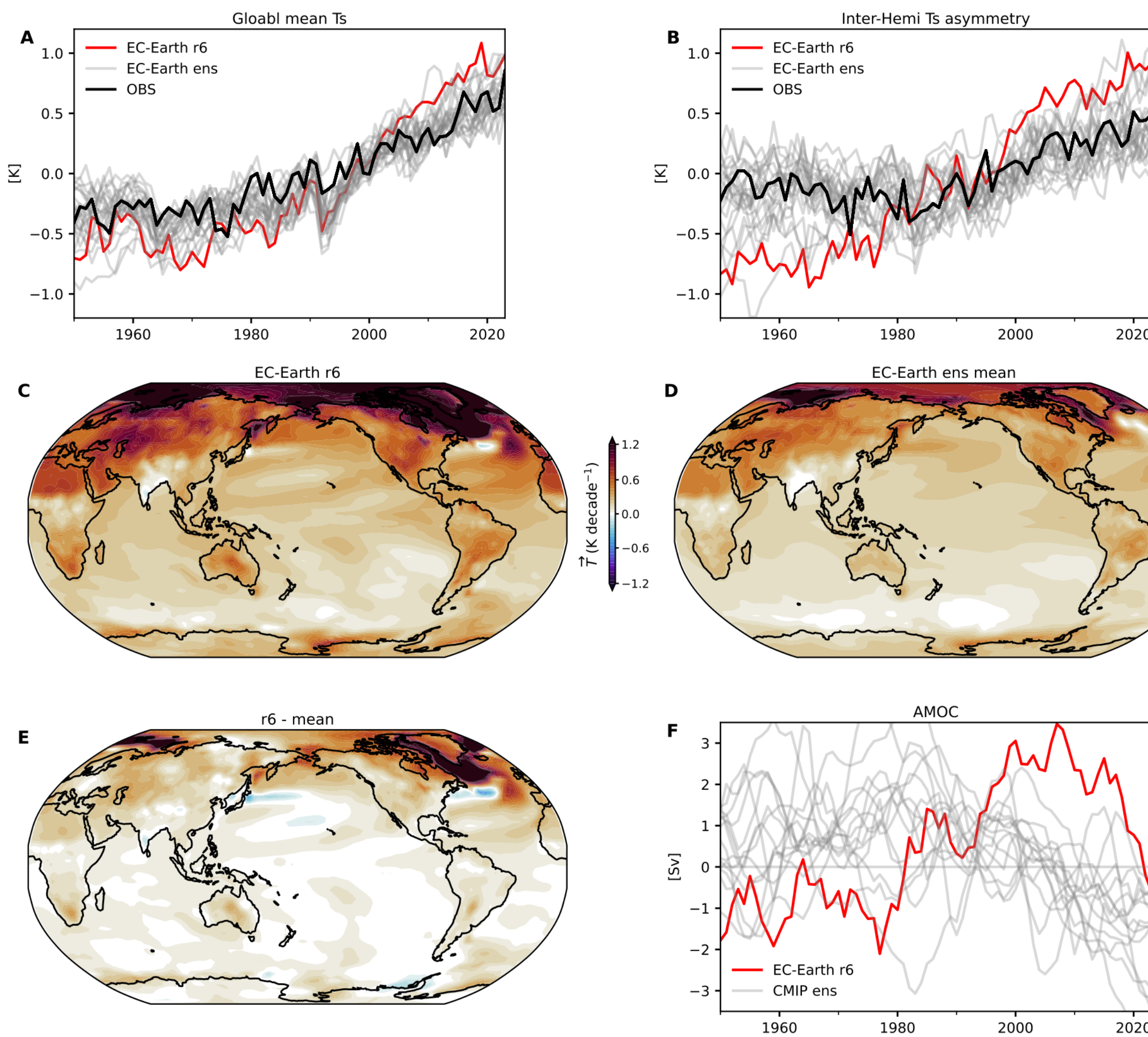

**Figure S1: Outlier behavior of EC-Earth3 r6.** (A) Time series of global mean Ts in EC-Earth r6 (red), other ensemble runs (grey), and observation (black thick). (B) Time series of interhemispheric asymmetry in Ts in EC-Earth r6 (red), other ensemble runs (grey), and observation (black thick). (C-E) Ts trend over 1980-2022 in EC-Earth r6 and in ensemble mean, and their difference. (F) Time series of AMOC strength in EC-Earth r6 (red) and other CMIP6 model realizations (grey).

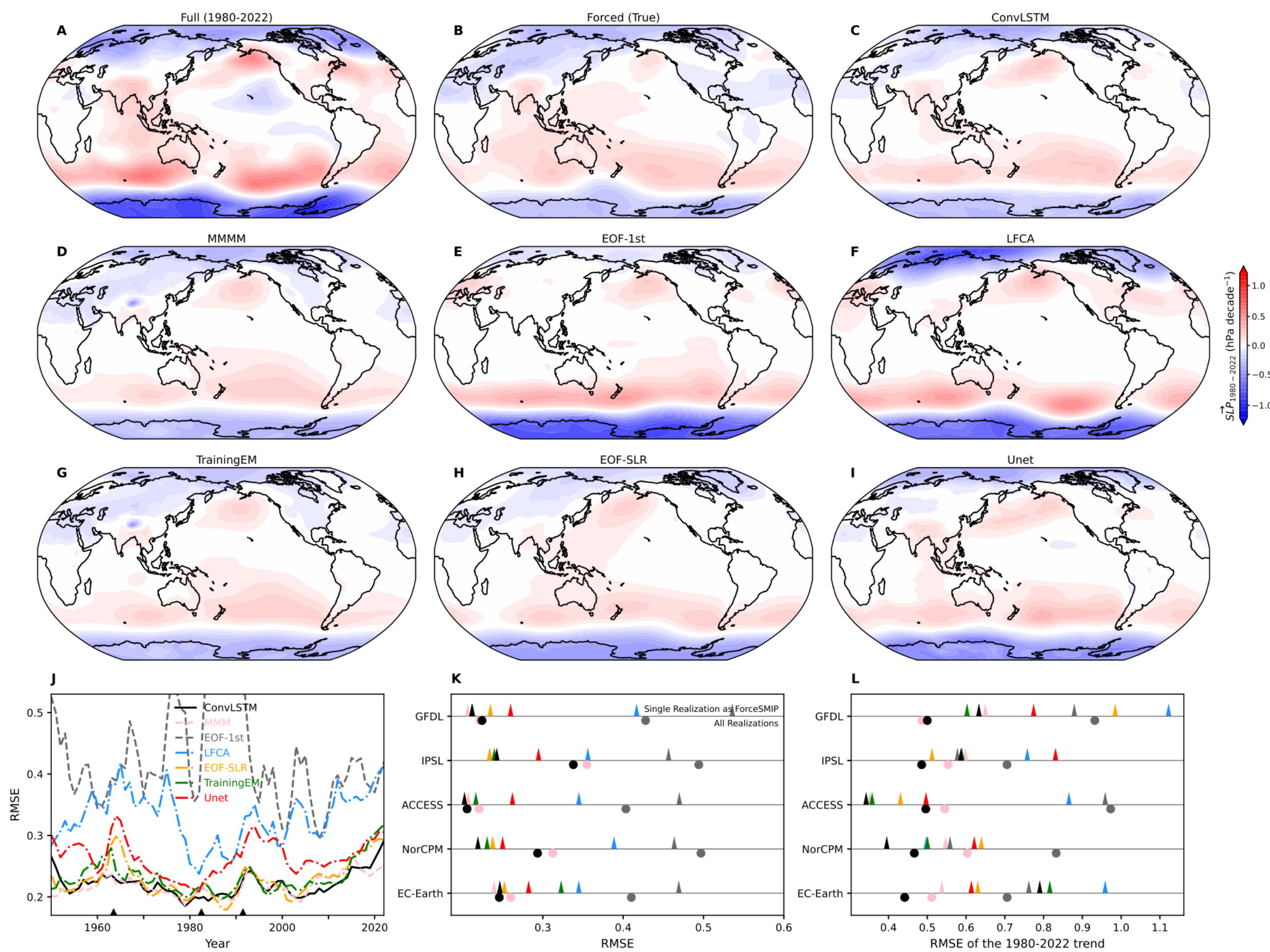

**Figure S2: Same as Fig. 1 but for sea-level pressure.**

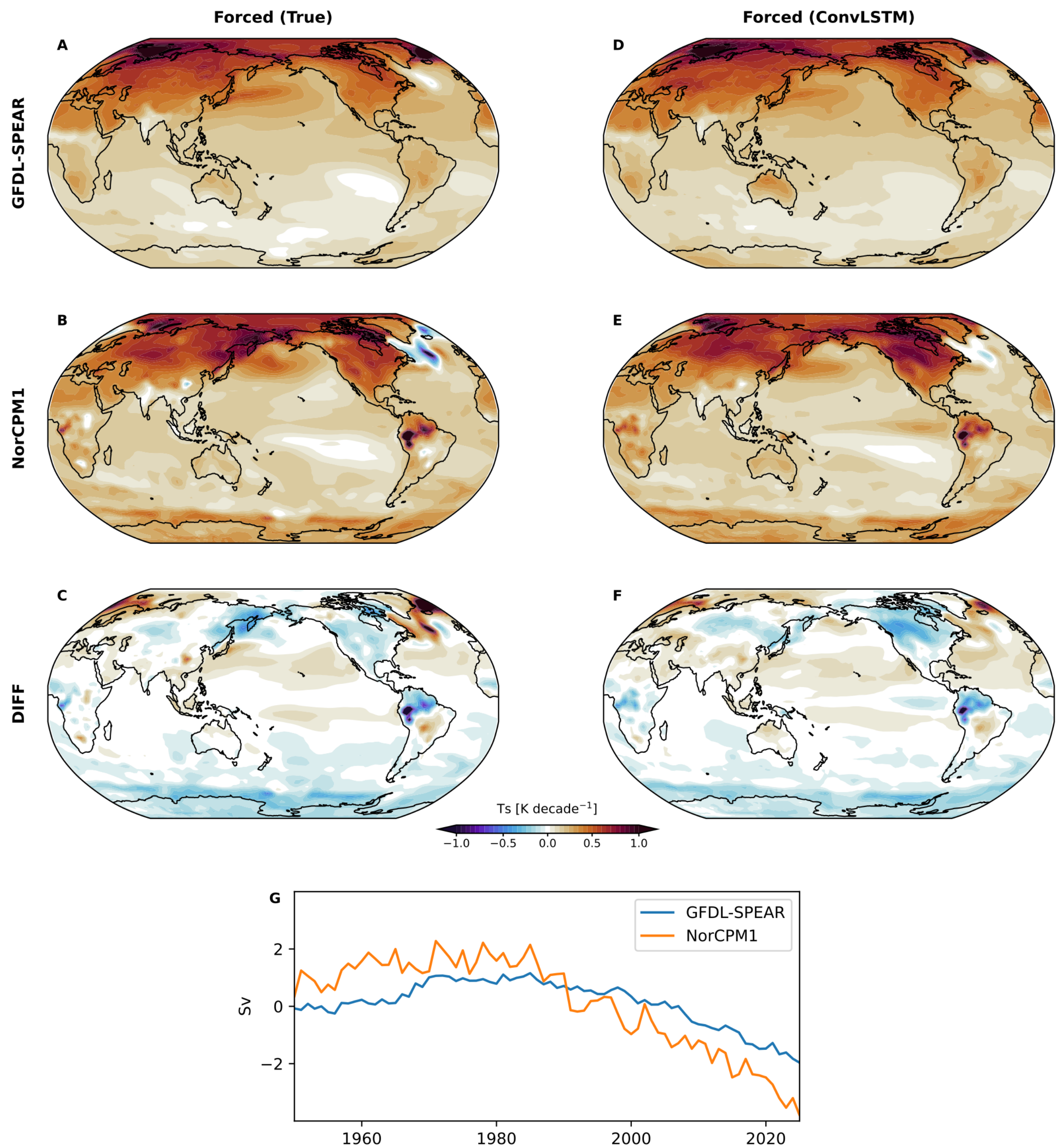

**Figure S3: Distinct forced response in SST between the evaluation cases 1E (GFDL-SPEAR r3) and 1J (NorCPM1 r4) captured by ConvLSTM.** (A-C) The true forced response for GFDL-SPEAR and NorCPM1, and their difference. (D-F) The forced response estimated by ConvLSTM based on GFDL-SPEAR r3 and NorCPM1 r4, and their difference. (G) The forced change in the AMOC strength in GFDL-SPEAR (blue) and NorCPM1 (orange).

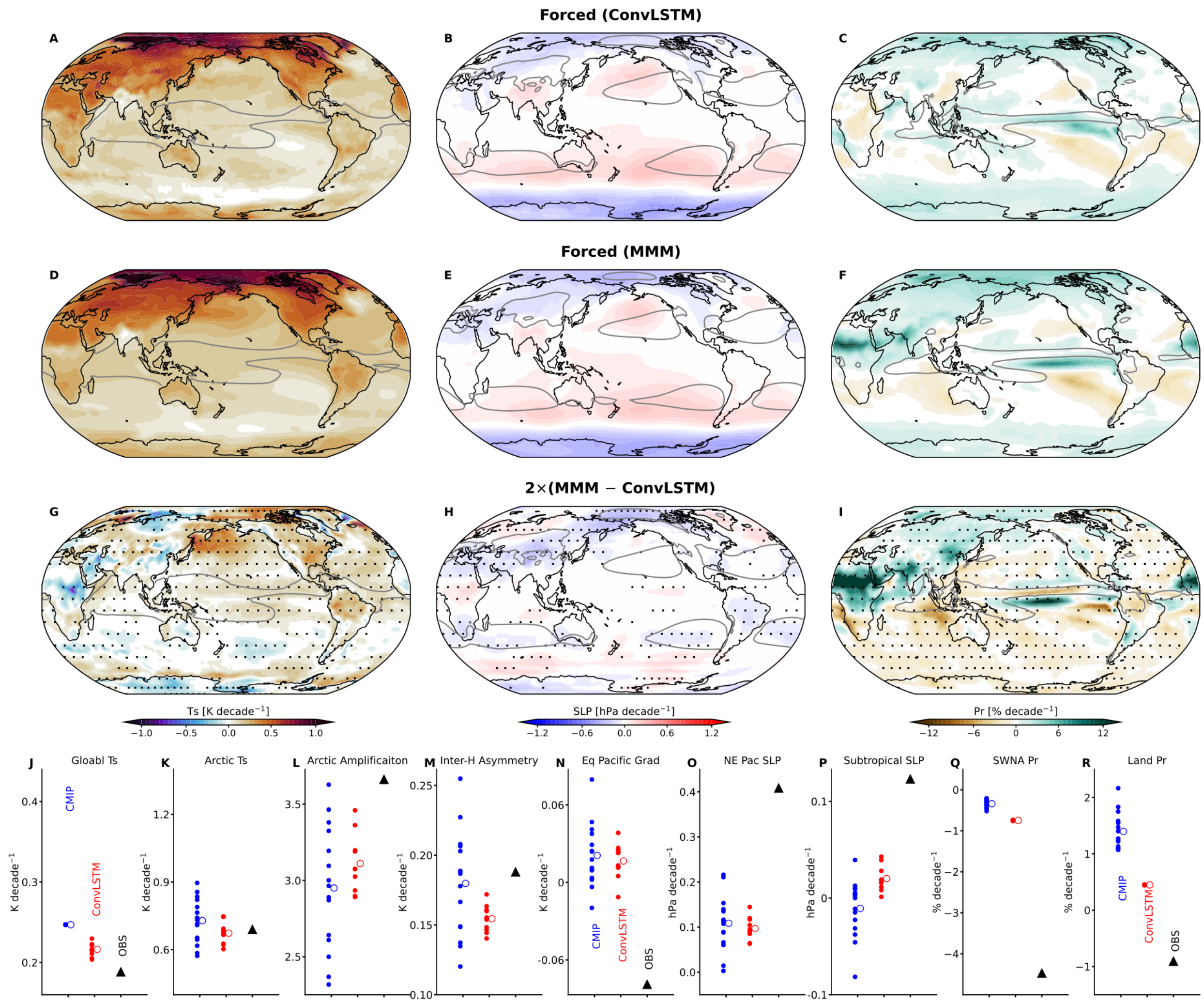

**Figure S4: Same as Fig. 7 but with the forced response estimated by climate models first normalized by their respective global mean warming rates and then averaged to obtain the multi-model mean (MMM).**

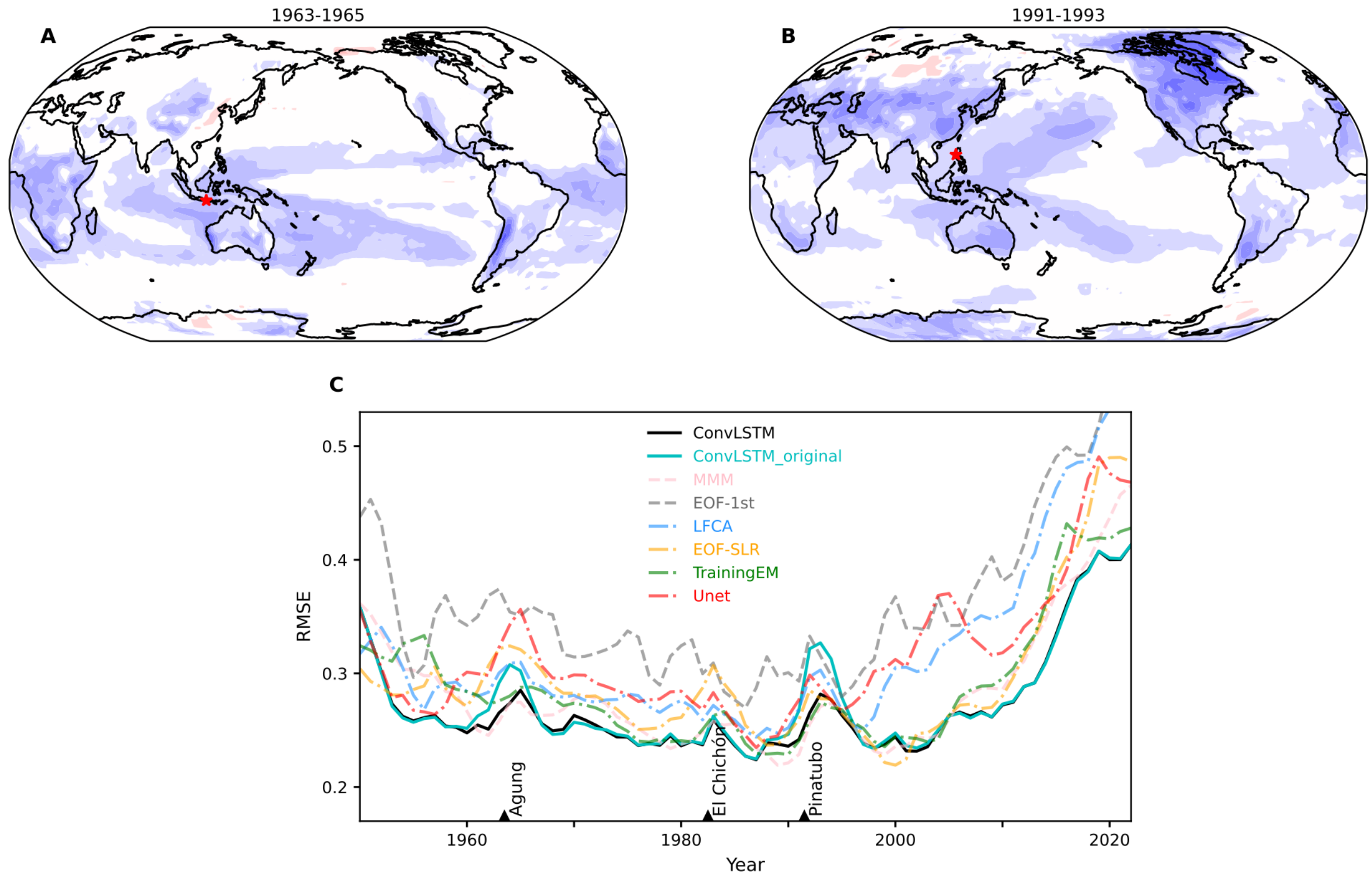

**Figure S5: Robust bias in the original estimate of ConvLSTM associated with volcano eruptions.** (A,B) Robust bias in the forced Ts in the original estimate of ConvLSTM associated with the Agung (1963-1965) and Pinatubo (1991-1993) Eruption, respectively. (C) Time series of the global mean RMSE of the forced Ts in various methods.

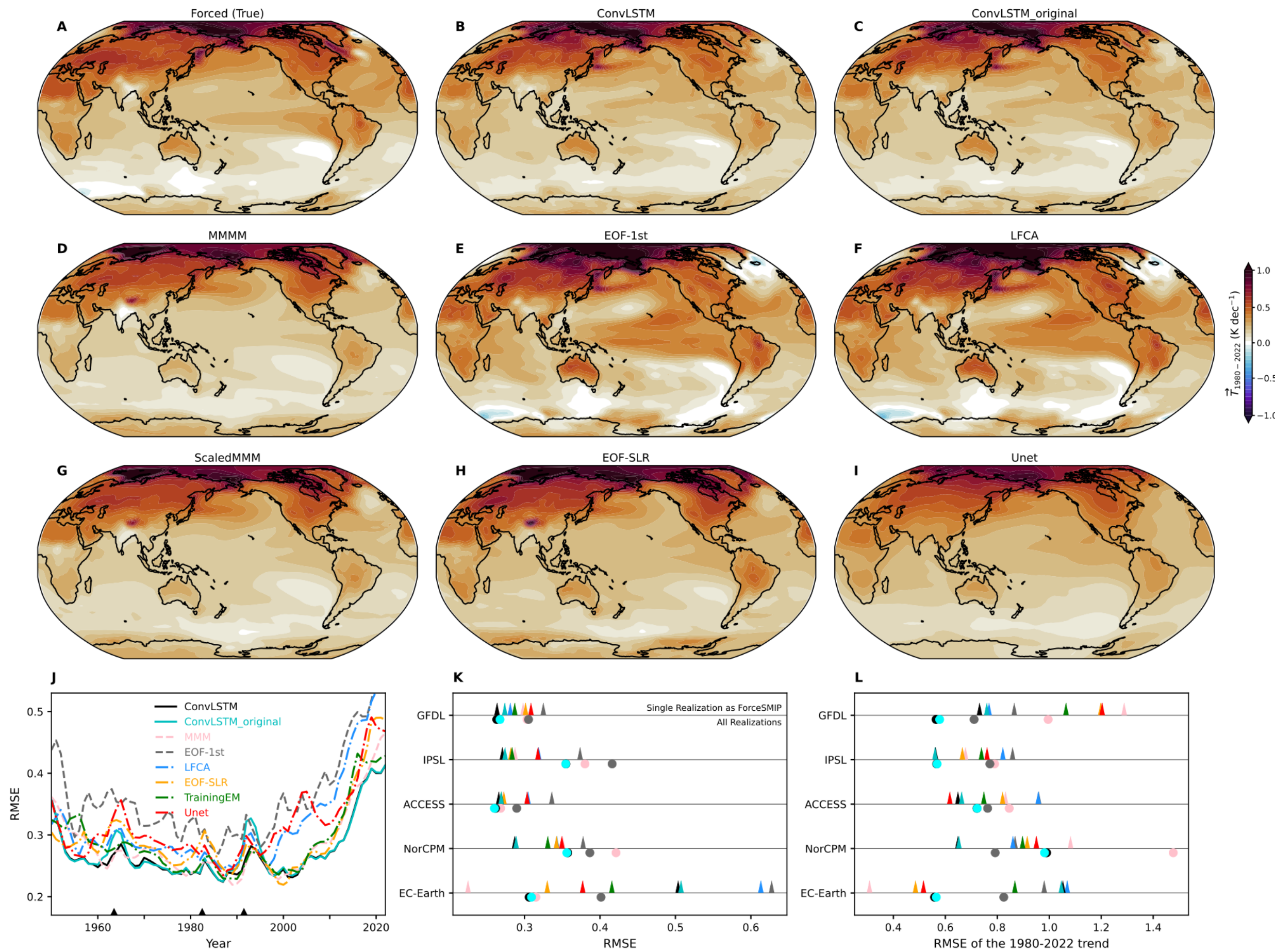

**Figure S6: Same as Fig. 2 but with the result of the original estimate of ConvLSTM included.**